\documentstyle[epsfig]{mn}

\newif\ifAMStwofonts

\ifoldfss
  \ifCUPmtlplainloaded \else
    \NewTextAlphabet{textbfit} {cmbxti10} {}
    \NewTextAlphabet{textbfss} {cmssbx10} {}
    \NewMathAlphabet{mathbfit} {cmbxti10} {} 
    \NewMathAlphabet{mathbfss} {cmssbx10} {} 
  \fi
  \ifAMStwofonts
    \ifCUPmtlplainloaded \else
      \NewSymbolFont{upmath} {eurm10}
      \NewSymbolFont{AMSa} {msam10}
      \NewMathSymbol{\upi}     {0}{upmath}{19}
      \NewMathSymbol{\umu}     {0}{upmath}{16}
      \NewMathSymbol{\upartial}{0}{upmath}{40}
      \NewMathSymbol{\leqslant}{3}{AMSa}{36}
      \NewMathSymbol{\geqslant}{3}{AMSa}{3E}

       \let\le=\leqslant
       \let\ge=\geqslant
    \fi
  \fi
\fi 

\ifnfssone
  \newmathalphabet{\mathit}
  \addtoversion{normal}{\mathit}{cmr}{m}{it}
  \addtoversion{bold}{\mathit}{cmr}{bx}{it}
  \newmathalphabet{\mathbfit} 
  \addtoversion{normal}{\mathbfit}{cmr}{bx}{it}
  \addtoversion{bold}{\mathbfit}{cmr}{bx}{it}
  \newmathalphabet{\mathbfss} 
  \addtoversion{normal}{\mathbfss}{cmss}{bx}{n}
  \addtoversion{bold}{\mathbfss}{cmss}{bx}{n}
  \ifAMStwofonts
    \ifCUPmtlplainloaded \else
and
your
      \UseAMStwoboldmath
      \makeatletter
      \new@mathgroup\upmath@group
      \define@mathgroup\mv@normal\upmath@group{eur}{m}{n}
      \define@mathgroup\mv@bold\upmath@group{eur}{b}{n}
      \edef\UPM{\hexnumber\upmath@group}
      \new@mathgroup\amsa@group
      \define@mathgroup\mv@normal\amsa@group{msa}{m}{n}
      \define@mathgroup\mv@bold\amsa@group{msa}{m}{n}
      \edef\AMSa{\hexnumber\amsa@group}
      \makeatother
      \mathchardef\upi="0\UPM19
      \mathchardef\umu="0\UPM16
      \mathchardef\upartial="0\UPM40
      \mathchardef\leqslant="3\AMSa36
      \mathchardef\geqslant="3\AMSa3E

       \let\le=\leqslant
       \let\ge=\geqslant
    \fi
  \fi
\fi 

\ifnfsstwo
  \DeclareMathAlphabet{\mathbfit}{OT1}{cmr}{bx}{it}
  \SetMathAlphabet\mathbfit{bold}{OT1}{cmr}{bx}{it}
  \DeclareMathAlphabet{\mathbfss}{OT1}{cmss}{bx}{n}
  \SetMathAlphabet\mathbfss{bold}{OT1}{cmss}{bx}{n}
  \ifAMStwofonts
    \ifCUPmtlplainloaded \else
      \DeclareSymbolFont{UPM}{U}{eur}{m}{n}
      \SetSymbolFont{UPM}{bold}{U}{eur}{b}{n}
      \DeclareSymbolFont{AMSa}{U}{msa}{m}{n}
      \DeclareMathSymbol{\upi}{0}{UPM}{"19}
      \DeclareMathSymbol{\umu}{0}{UPM}{"16}
      \DeclareMathSymbol{\upartial}{0}{UPM}{"40}
      \DeclareMathSymbol{\leqslant}{3}{AMSa}{"36}
      \DeclareMathSymbol{\geqslant}{3}{AMSa}{"3E}

       \let\le=\leqslant
       \let\ge=\geqslant
    \fi
  \fi
\fi 

\ifCUPmtlplainloaded \else
  \ifAMStwofonts \else 
    \def\upi{\pi}
    \def\umu{\mu}
    \def\upartial{\partial}
  \fi
\fi

\title{Cosmic metal production and the mean metallicity of the Universe}
\author[F. Calura, F. Matteucci]
       {F. Calura$^{1}$\thanks{E-mail: fcalura@ts.astro.it}, 
        F. Matteucci$^{1}$\\
        (1) Dipartimento di Astronomia-Universit\'a di Trieste, Via G.
B. Tiepolo
	11, 34131 Trieste, Italy\\
	 }
	
\date{Accepted for publication 22/01/04}

\pagerange{\pageref{firstpage}--\pageref{lastpage}}
\pubyear{2004}

\begin{document}

\maketitle

\label{firstpage}
\begin{abstract}
By means of detailed chemo-photometric models for elliptical, spiral and irregular galaxies,  
we evaluate the cosmic history of the production of chemical elements as well as the metal mass density of 
the present-day universe. 
In particular, we study the production rates of some of the most common 
chemical species (O, Mg, N, Si, Fe, Zn), detected both in local stars, galactic and extragalactic 
HII regions and high redshift objects. Such a study allows us to compute in detail the fraction of each element   
locked up in stars, interstellar gas and intergalactic medium.  
We then calculate the mean metal abundances for galaxies of different morphological types, along with the average metallicity of 
galactic matter in the universe (stars, gas and intergalactic medium). For the average metallicity of galaxies in the local universe, we find $<Z>_{gal}= 0.0175$, i.e. 
close to the solar value. We find the main metal production in spheroids (ellipticals and bulges) to occur at very early times,  
implying an early peak in the metal production and a subsequent decrease. On the other hand, the metal 
production in spirals and irregulars is always increasing with time. We find that the average $[O/Fe]_{*,E}$ ratio in stars 
in spheroids should be $+0.4$ dex, whereas the average $[O/Fe]_{gas,E}$ ratio in the gas should be $-0.33$ dex, 
due to the large amount of Fe produced in these systems by SNae Ia after star formation has stopped. The same quantities for spirals are 
$\sim +0.1$ dex for the stellar component and almost solar ($\sim +0.01$ dex) for the gas component. 
We suggest that a Salpeter-like IMF is the best candidate for the universal IMF since it allows us to reproduce the majority 
of observational constraints. 
We perform a self-consistent census of the baryons and metals in the local universe finding that, while the vast 
majority of the baryons lies outside galaxies in the inter-galactic medium (IGM), $52 \%$ of the metals (with the exception 
of the Fe-peak elements) is locked up in stars and in the interstellar medium.
We estimate indirectly the amount of baryons which resides in the IGM  
and we derive its mean Fe abundance, finding a value of $X_{Fe,IGM}=0.05 X_{Fe,\odot}$. We believe that this estimate is uncertain 
by a factor of $\sim 2$, owing to the normalization of the local luminosity function.   
This means that the Fe abundance of $0.3$ solar inferred from X-ray observations of the hot intra-cluster medium (ICM)  
is higher than the average Fe abundance of the inter-galactic gas in the field.  
\end{abstract} 


\section{Introduction}
In the history of the universe, there have been two fundamental steps 
in the production of the chemical elements. 
The first has been the epoch of primordial nucleosynthesis, occurred 
immediately after the Big Bang, which has led to the production of
Deuterium, $^{3}$He, $^{4}$He and small traces of $^{7}$Li. 
During the second epoch, practically all the metals have been synthesized 
by stars within galaxies, starting 
soon after the recombination era up to the present time.
The metals can have very different histories, strictly
linked 
to the processes governing the stellar nucleosynthesis and the star formation history, fundamental 
in the study of how galactic structures have evolved once
assembled. 
Part of the metals remains locked up in long living stars and remnants, 
part is restored into the galactic inter-stellar medium (ISM) 
through supernova (SN) explosions and stellar winds while 
another fraction is expelled in the inter-galactic medium (IGM) through 
galactic winds and outflows (for a comprehensive review of element production through the
 cosmic ages, see Pettini 2002).\\
The observational study of metal abundances in various components of 
the universe has experienced a considerable development in the last few
decades. 
Thanks to the construction of high-performance telescopes, it has been 
possible to measure the abundances of a wide range of chemical elements, both in the local and
in the high-redshift universe. All these studies indicate a strong chemical evolution.  
The theoretical investigation of these aspects is performed by means of cosmic chemical evolution models 
(Pei \& Fall 1995, 
Edmunds \& Phillips 1997, Malaney \& Chaboyer 1996, Pei, Fall \& 
Hauser 1999, Cen \& Ostriker 1999, Sadat Guiderdoni \& Silk 2001, Mathlin et al. 2001), very fruitful in 
tracing the average properties of the universe but unsuited to 
study in detail the different contributions to the overall metal content of the universe from 
various galactic types, namely spheroids, spiral discs and irregulars. 
Furthermore, with the exception of Malaney \& Chaboyer (1996),
all of these models are generally developed 
to account for the total metallicity Z, which can be studied  
by means of simplifying assumptions on the stellar lifetimes such as the  
instantaneous recycling approximation (IRA, Tinsley 1980, Matteucci 2001), 
but do not allow to predict the
evolution of the abundances of single elements, generally produced on very different timescales.\\
Calura \& Matteucci (2003, hereinafter CM03) have developed a series of chemical and 
spectro-photometric models for elliptical, spiral and irregular galaxies. 
These models have been primarily used to study the evolution of the luminous matter 
in the universe and the contributions that spheroids, spiral discs and irregulars bring to the 
overall cosmic star formation. Beside this, such models were used to predict the 
early phases of the evolution of the total metal content in the universe, in 
particular to investigate the issue of the missing metals, but no detailed prediction about the single chemical elements 
were presented. In this paper, by means of the chemical evolution models developed by CM03, 
we aim at reconstructing the cosmic evolution of single elements 
in the
total stellar and gaseous components in galaxies of different morphological types. 
In particular, we study the 
evolution in time of several chemical elements commonly detected in local
stars and HII regions as well as in high redshift galaxies, such as O, Mg, N, Si,
Fe, Zn.
We also study the cosmic and stellar production of $^4$He  
and, by means of $^4$He abundances measured in local 
blue compact dwarf (BCD) galaxies and those inferred for the Galactic Bulge, 
we try to give an independent estimate of the primordial $^4$He abundance $Y_{p}$.  
We then calculate the mean abundances of the above mentioned elements 
in spheroids, spiral discs and irregulars, along with the average metallicity of 
galactic matter in the universe.  
We attempt an indirect estimate of the average metallicity of the 
IGM (measured by the Fe abundance) and we perform a 
self-consistent census of the baryons and metals in the 
universe at the present time. Our results are compared with  
both observational and theoretical appraisals by various authors.\\
In section 2 we briefly describe the theory on which our methods 
are based and their physical motivations, in section 3 we present 
our results and in  section 4 we draw the conclusions. 
Throughout the paper, to facilitate comparison with other published results we assume an Einstein-De Sitter
cosmology ($\Omega_{m}=1$, $\Omega_{\Lambda}=0$) and $H_{0}=50 Km/s Mpc$.

\section{The chemical evolution models}
Chemical evolution models allow us to follow in detail the evolution of the abundances of several chemical
species, starting from the matter reprocessed 
by the stars and restored into the ISM through stellar winds and supernova
explosions. 
Detailed descriptions of the chemical evolution models can be found in Matteucci
\& Tornamb\'{e} (1987) and Matteucci (1994) for elliptical galaxies, 
Chiappini et al. (1997, 2001) for the spirals and 
Bradamante et al. (1998) for irregular galaxies.
We differentiate the galaxy types into ellipticals, spirals and irregulars. 
We assume that the category of galactic bulges is naturally included in the one of elliptical galaxies. 
Our assumption is motivated by the fact that they have very similar features: for instance, 
both are dominated by old stellar populations and respect the same fundamental plane (Binney \&
Merrifield 1998). 
This certainly indicates that they are likely to have a common origin, i.e. both are likely to have
formed on very short timescales and a long time ago.\\
In our picture, spheroids form as a result of the rapid collapse of a homogeneous sphere of
primordial gas where star formation is taking place at the same time as the collapse proceeds. 
Star formation is assumed to halt as the energy of the ISM, heated by stellar winds and SN explosions,
balances the binding energy of the gas. At this time a galactic wind occurs, sweeping away almost all of  
the residual gas. By means of the galactic wind, ellipticals enrich the IGM with metals.\\ 
For spiral galaxies, the adopted model is calibrated in order to reproduce a large set of observational 
constraints for the Milky Way galaxy (Chiappini et al. 2001). The Galactic disc is approximated by several independent rings, 
2 kpc wide, without exchange of matter between them. In our picture, 
spiral galaxies are assumed to form as a result of two main infall episodes.  
During the first episode, the halo and the thick disc are formed.
During the second episode, a slower infall
of external gas forms the thin disc with the gas accumulating faster in the inner than in the outer
region ("inside-out" scenario, Matteucci \& Fran\c cois 1989). The process of disc formation is much longer 
than the halo 
and bulge formation, with time scales varying from $\sim2$ Gyr in the inner disc to $\sim8$ Gyr in the solar region
and up to $15-20$ Gyr in the outer disc.\\
There are two differences between the model adopted here and the one 
by Chiappini et al. 2001: the choice of the IMF and the 
lack of the star formation threshold. 
The first is due to the fact that our aim is to test the hypothesis of a universal IMF. 
To this purpose, we have 
investigated the results obtained either with a Salpeter or a Scalo IMF, 
both extended to all the morphological types. 
Secondly, the elimination 
of the star formation threshold is motivated by the fact that its effects 
are appreciable only on 
small scales, i.e. in the chemical evolution of the solar vicinity and of small galactic regions, whereas 
our aim is to study the element production in galactic discs on global scales. 
Finally, irregular dwarf galaxies are assumed to assemble from merging of protogalactic small clouds 
of primordial chemical composition, until masses in the range $\sim 10^{8} - 6 \times 10^{9}M_{\odot}$ are accumulated, 
and to produce stars at a lower rate than spirals. 
Our typical irregular has a mass of $6 \times 10^{9}M_{\odot}$, i.e. 
comparable to the one of the Large Magellanic Cloud (Russel \& Dopita 1992).\\
Let $G_{i}$ be the fractional mass of the element $i$ in the gas
within a galaxy, its temporal evolution is described by the basic equation:\\
\begin{equation}
\dot{G_{i}}=-\psi(t)X_{i}(t) + R_{i}(t) + (\dot{G_{i}})_{inf} - 
(\dot{G_{i}})_{out}\\
\end{equation} 
where $G_{i}(t)=M_{g}(t)X_{i}(t)/M_{tot}$ is the gas mass in the form of an
element $i$ normalized to a total initial mass $M_{tot}$. The quantity $X_{i}(t)=
G_{i}(t)/G(t)$ represents the abundance in mass of an element $i$, with
the summation over all elements in the gas mixture being equal to unity.
The quantity $G(t)= M_{g}(t)/M_{tot}$ is the total fractional mass of gas
present in the galaxy at time t.
$\psi(t)$ is the instantaneous star formation rate (SFR), namely the fractional amount
of gas turning into stars per unit time; $R_{i}(t)$ represents the returned
fraction of matter in the form of an element $i$ that the stars eject into the ISM through stellar winds and 
SN explosions; this term contains all the prescriptions regarding the stellar yields and
the SN progenitor models.
The two terms 
$(\dot{G_{i}})_{inf}$ and $(\dot{G_{i}})_{out}$ account for the infalling
external gas from the IGM and for the outflow, occurring
by means of SN driven galactic winds, respectively. 
The main feature characterizing a particular morphological galactic type is
represented by the prescription adopted for the star formation history.\\ 
In the case of elliptical and irregular  
galaxies the SFR $\psi(t)$ (in $Gyr^{-1}$) has a simple form and is given by:

\begin{equation}
\psi(t) = \nu G(t) 
\end{equation}

The quantity $\nu$ is the efficiency of star formation, namely the inverse of
the typical time scale for star formation and for ellipticals and bulges is assumed to 
be $\sim 11 $ Gyr$^{-1}$. In the case of ellipticals, $\nu$ is
assumed to drop to zero at the onset of a galactic wind, which develops as the
thermal energy of the gas heated by supernova explosions exceeds the binding
energy of the gas (Arimoto \& Yoshii 1987, Matteucci \& Tornamb\'{e} 1987). 
This quantity is strongly
influenced by assumptions concerning the presence and distribution of dark
matter (Matteucci 1992); 
for the model adopted here a diffuse 
($R_e/R_d$=0.1, where
$R_e$ is the effective radius of the galaxy and $R_d$ is the radius 
of the dark matter core) but 
massive ($M_{dark}/M_{Lum}=10$) dark halo has 
been assumed.\\
In the case of irregular galaxies we have assumed a continuous star formation rate always expressed as in (2), but
characterized by an efficiency lower than the one adopted for ellipticals, in particular $0.01$ Gyr$^{-1}$\\
In the case of spiral galaxies, the SFR expression (Matteucci \& Fran\c cois 1989) is:\\
\begin{equation}
\psi(r,t) = \nu [\frac{\sigma(r,t)}{\sigma(r_{\odot},t)}]^{2(k-1)} [\frac{\sigma(r,t_{Gal})}{\sigma(r,t)}]^{k-1}
G^{k}(r,t)
\end{equation}

where $\nu$ is the SF efficiency, $\sigma(r,t)$ is the total (gas + stars) surface mass density at a radius r and
time t, 
$\sigma(r_{\odot},t)$ is the total surface mass density in the solar region. 
For the gas density exponent $k$ a value of 1.5 has been assumed, in order to ensure 
a good fit to the observational constraints at the solar vicinity, in agreement with Kennicutt (1998).\\
The quantity $G(t)$ for galactic discs is the same as before but is expressed as a function 
of the surface densities ($G(t)=\sigma_{gas}/\sigma$). 
For the stellar yields, for massive stars (M $> 8 M_{\odot}$) 
we adopt nucleosynthesis 
prescriptions by 
Nomoto et al. (1997a), which present an uptodate and 
more extensive grid of models relative to Thielemann, Nomoto \& 
Hashimoto (1996). These yields are calculated as functions of the mass 
of the He cores. In order to obtain the total ejected mass in the form of 
single elements, we compute the difference between the initial mass of the star 
and the mass of the He core ($M-M_{\alpha}$) and multiply this quantity by the 
abundance of each element originally present in the star at birth. 
This procedure is contained in the production matrix (Talbot \& Arnett 1971) which enters into the basic 
equations of chemical evolution (see eq. 6). A similar procedure was adopted by Thomas, Greggio \& Bender (1998). 
We adopt the yields by van den Hoeck \& Groenewegen (1997) for
low and intermediate mass stars ($0.8 \le M/M_{\odot} \le 8$) and 
Nomoto et al. (1997b) for type I a SNe. For the evolution of Zn and Ni we adopted
the nucleosynthesis prescriptions of Matteucci et al. (1993).\\
The IMF has been assumed to be constant in space and time. We have compared 
results calculated with two different IMFs, i. e. the Salpeter (1955) and the Scalo (1986), and we have 
also investigated the effects of the assumption of a flatter IMF.\\
The progenitors of type Ia SNe are assumed to be single degenerate systems according to the 
prescriptions of the Matteucci \& Recchi (2001) best model.\\
We have calculated the comoving densities of element production by 
normalizing the galaxy densities and the fractions of the various morphological types 
according to the local B-band luminosity function by Marzke et al. (1998), and assumed a scenario of pure 
luminosity evolution, namely that galaxies evolve 
only in luminosity and not in number. Such a picture can account for many observables, such as the evolution 
of the galaxy luminosity density in various bands and the cosmic supernova rates (CM03).
 To calculate the B band luminosities and evolutionary corrections, we have used the 
spectro-photometric code by Jimenez et al. (1998), which allows us to compute self-consistently the 
photometric evolution of composite stellar populations taking into account the metallicity of the gas 
out of which the stars form at any time (see CM03 for a more detailed description).\\

\section{Results}

\subsection{Evolution of $^4$He and metal production rates}
The total production rate for the $i-$th element $\dot{\rho_{i}}$ is given by:\\
\begin{equation}
\dot{\rho_{i}}(M_{\odot} \, yr^{-1} \, Mpc^{-3})=\sum_{k} \rho_{B,k}(z) (M/L)_{B,k}(z) \gamma_{i \, k}(z) \\
\end{equation} 
where $\rho_{B,k}$ is the B band luminosity density for the $k-$th morphological type and 
is the result of the the integral over 
all luminosities of the B-band luminosity function (LF) $\Phi(L_{B})$:\\
\begin{equation}
\rho_{B}= \int{\Phi(L_{B})\, (L_{B}/L^{\ast}_{B})\, dL_{B}} \\
\end{equation}
In the present work, we have adopted the LF observed by Marzke et al. (1998) for all the 
different morphological types. 
At redshift other than zero, we have calculated the LFs 
by applying the evolutionary corrections calculated by means of the spectrophotometric model 
(this procedure is described in detail in Calura \& Matteucci 2003). 
$(M/L)_{B,k}$ are the B mass-to-light ratios for galaxies of the $k-th$ morphological type. 
For each galactic type, the evolution of the stellar, gaseous and total mass is a prediction of the 
chemical evolution models, 
whereas the evolution of the B luminosity $L_{B}$ is calculated by means of the spectro-photometric 
code. 
$\gamma_{i \, k}$ represents 
the rate of production of the $i-$th element for the galaxy of $k-$th morphological type, and at a given 
time t and is:\\
\begin{equation}
\gamma_{i \, k}(t) =  \int_{m(t)}^{M_{max}} \psi_{k}(t-\tau_{m}) \, R_{mi}(t-\tau_{m}) \, \phi(m) dm  \\
\end{equation} 
$m(t)$ is the minimum mass of the stars dying at the time t, $M_{max}=100 M_{\odot}$ is the upper mass limit 
considered in the IMF. $\tau_{m}$ is the lifetime of the stars of mass $m$, $\psi_{k}$ is the SFR and  $R_{mi}$ is the fraction  
of the mass of a star of mass $m$ which is ejected back into 
the ISM in the form of species i. 
In order to compute $R_{mi}$ we adopt the Talbot \& Arnett (1971) production 
matrix (see Matteucci 2001 for details). 
Finally, the conversion between the cosmic time $t$ into redshift is performed on the basis of the 
adopted cosmological model. 
Figure 1 shows the calculated total production rates for all the elements considered in the present work, 
along with the single contributions from spheroids, spiral discs and irregulars in the case of a Salpeter 
IMF and having assumed that all galaxies started forming stars at $z_{f}=5$. We stress that all our results are 
fairly insensitive to the choice of $z_{f}$. For all the elements, our predictions 
indicate a strong peak in the production rate at early times due to spheroids, consequent to the intense
star-burst which they experience soon after their formation. Later on, the production rate in 
spheroids diminishes and at redshift $\sim 2$ the main sites of element production become the spiral 
discs. The lowest star formation rates characterize the irregular systems, whose contribution to the total 
element production is in general negligible. Note that at $z \sim 0$ the spheroid production rates for all the elements 
falls at a level comparable or lower than the one of irregulars. 
On the other hand, from the beginning down to $z \sim 2$ the spheroids continue to be intense producers (comparable to discs) 
of elements synthesized on long timescales ($\ge 1 Gyr$) by intermediate mass stars such as Fe and Zn, 
restored into the ISM via type Ia SNe. 
In figure 2 we show the total production rates and the single contributions by spheroids, discs and 
irregulars in the case of a Scalo IMF. The curves have the same behaviours as in the Salpeter case, 
with the element production dominated by spheroids at early times and by discs successively.  
In figure 3 we show a comparison among the total element production rates calculated with the 
Salpeter, the Scalo and the Arimoto-Yoshii (1987, AY) IMFs, this latter characterized by a slope of 
$x = 0.95$, hence much 
richer in high mass stars. As expected, the highest amounts of $^4$He and metals are produced with an AY IMF, 
whereas the lowest amount is produced with the Scalo IMF.
In the next section we discuss the most appropriate IMF  
for the whole galaxy population by comparing the integrated production rates and the mean abundances 
calculated for various galaxies. 

\begin{figure*}
\centering
\vspace{0.001cm}
\epsfig{file=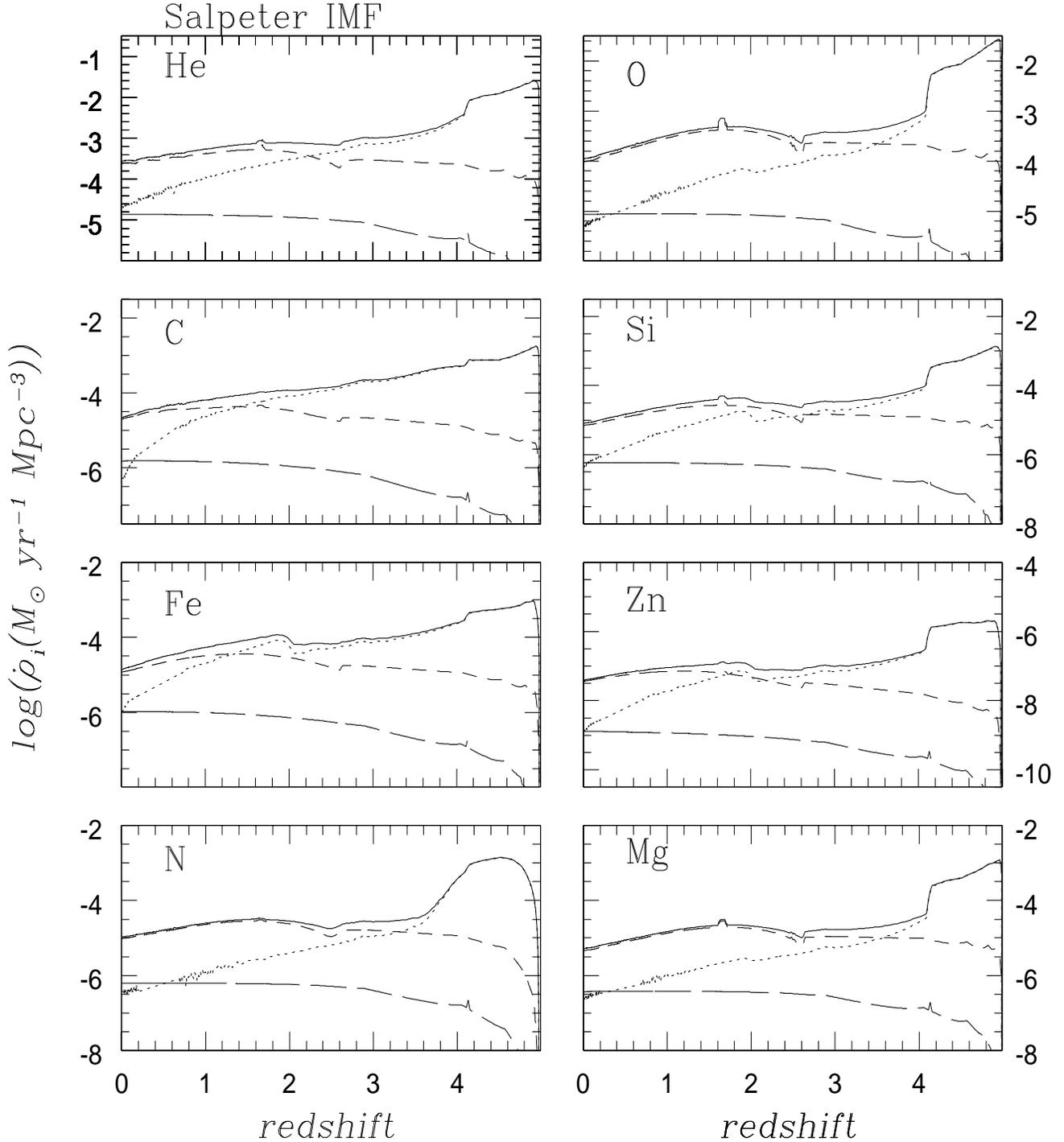,height=20cm,width=18cm}
\caption[]{Predicted production rates vs redshift for $^4$He, O, C, Si, Fe, Zn, N, and Mg assuming a Salpeter IMF. 
\emph{Short-dashed lines}: contribution by spiral galaxies; 
\emph{dotted lines}: contribution by spheroids;  
\emph{long-dashed lines}: contribution by irregulars. \emph{Solid lines}: total production rate densities.   }	
\end{figure*}
\begin{figure*}
\centering
\vspace{0.001cm}
\epsfig{file=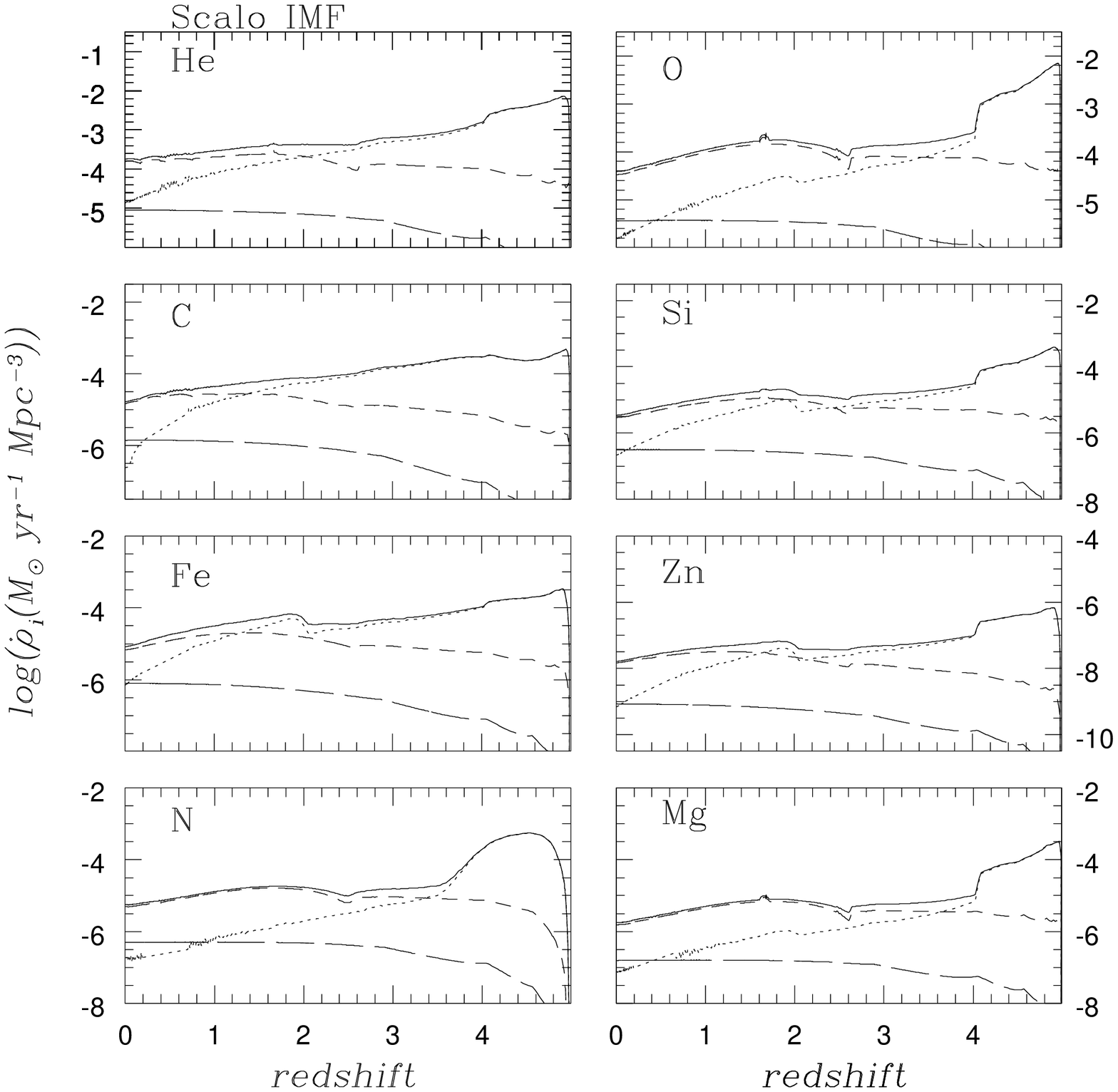,height=20cm,width=18cm}
\caption[]{Predicted production-rate densities vs redshift for $^4$He, O, C, Si, Fe, Zn, N, and Mg assuming a Scalo IMF. 
\emph{Short-dashed lines}: contribution by spiral galaxies; 
\emph{dotted lines}: contribution by spheroids;  
\emph{long-dashed lines}: contribution by irregulars. \emph{Solid lines}: total production rate densities. }	
\end{figure*}
\begin{figure*}
\centering
\vspace{0.001cm}
\epsfig{file=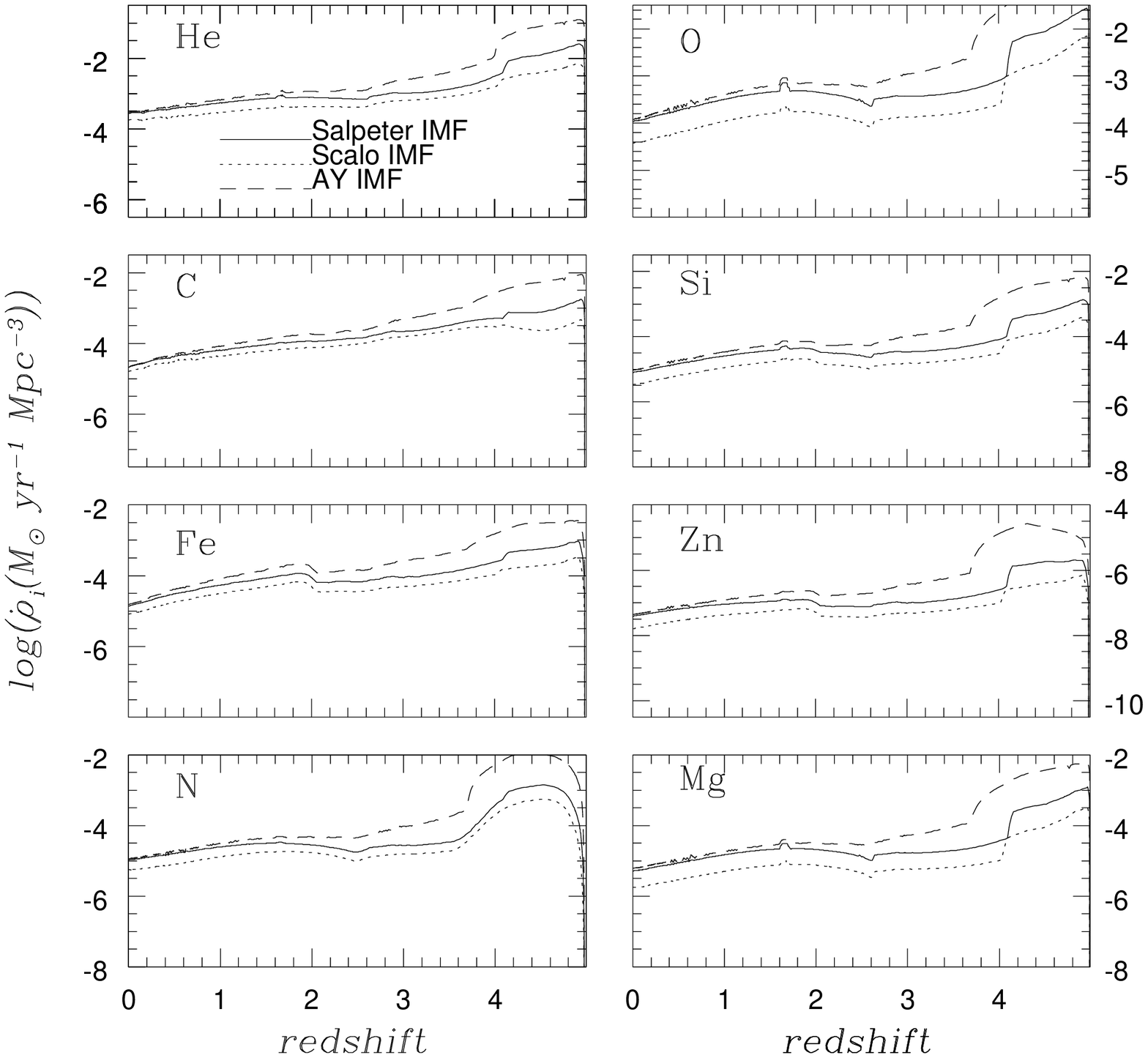,height=20cm,width=18cm}
\caption[]{Predicted total production rates vs redshift for $^4$He, O, C, Si, Fe, Zn, N, and Mg for different choices of the IMF.
\emph{Solid lines}: Salpeter IMF; 
\emph{dotted lines}: Scalo IMF;  
\emph{long-dashed lines}: AY IMF. }
\end{figure*}

\subsection{$^4$He and metal comoving densities}

At the present time, the \underline{total} comoving mass density in the form of a given element (labelled $i$) 
produced in the $k-$th galaxy type  
$\rho_{i,k,Tot} (M_{\odot} Mpc^{-3})$ is given by 
the integral of the production rate $\dot{\rho}_{k,i}$ calculated
over a Hubble time $T_{H}$:\\
\begin{equation} 
\rho_{i,k,Tot} = \int_{0}^{T_{H}} \dot{\rho}_{i,k}(t)dt \\
\end{equation}
In this case we assume $T_{H}=12$ Gyr, i.e. the age of all galaxies having assumed a redshift of 
galaxy formation $z_{f}=5$ for the cosmology adopted here. 
\underline{We divide the metal content of the universe} 
\underline{into three phases:}\underline{ the metals locked up in the stars,} \underline{in 
the ISM gas and in the IGM. }
At each instant, we can predict the comoving density of metals contained in each phase. 
To calculate the comoving mass density for the element $i$ in stars, ISM gas and IGM, 
first we need to calculate  the average stellar and ISM abundances. These quantities are direct outputs of our chemical evolution models  
and will be discussed in section 3.3, where an estimate of the average metallicity in galaxies will be provided. 

Let us define $<X_{i}>_{*,k}$ and $<X_{i}>_{g,k}$ as the average 
abundances by mass in stars and ISM gas, respectively, for the element $i$ and the $k$ galaxy type. 
The stellar comoving density of the element $i$ in the $k-th$ galaxy type is given by:\\
\begin{equation}
\rho_{i,k,*}=<X_{i}>_{*,k} \cdot \rho_{*,k} 
\end{equation}
whereas the comoving density of the element $i$ in ISM gas in the $k-th$ galaxy is given by:\\
\begin{equation}
\rho_{i,k,ISM}=<X_{i}>_{g,k} \cdot \rho_{g,k} 
\end{equation}
$\rho_{*,k}$ and $\rho_{g,k}$ are the mass densities of stars and ISM gas, respectively, 
and are  calculated as:
\begin{equation} 
\rho_{*,k}=\rho_{B,k} \cdot (M_{*}/L)_{B,k} \\
\end{equation}
and 
\begin{equation} 
\rho_{g,k}=\rho_{B,k} \cdot (M_{g}/L)_{B,k},\\ 
\end{equation}
where $\rho_{B,k}$ is the predicted B luminosity density, whereas $(M_{*}/L)_{B,k}$ and 
$(M_{g}/L)_{B,k}$ represent the predicted stellar and gas mass to light ratios, respectively, for the $k-$th galactic 
morphological type.\\
Here, we assume that only elliptical galaxies contribute to the chemical enrichment of the IGM through enriched 
galactic winds. 
The comoving density of a given element $i$ expelled by ellipticals (labelled E) into the IGM is given by:\\
\begin{equation} 
\rho_{i,E,IGM}= \rho_{i,E,Tot} - \rho_{i,E,*} - \rho_{i,E,ISM}
\end{equation} 

In Table 1 we show all the calculated comoving 
densities for various elements as produced by elliptical galaxies, in the case of a 
a Salpeter IMF. In the first column we show the symbols of the 
chemical elements, in columns 2-5 we report the present 
comoving mass density for the $i-$th element locked up in stars, ISM gas, IGM and the total comoving mass density. 
In Tables 2 and 3 we present the calculated comoving 
densities for various elements produced in spiral and irregular galaxies, respectively. 
In both tables, in the first column the symbols of the elements are shown, in the second and third columns we report the 
comoving densities of elements locked in stars and ISM gas, respectively, whereas in the fourth column the 
total comoving densities are shown. In fact, for these galaxies we did not take into account 
the presence of galactic winds. 
Tables 4, 5 and 6 show the same as Tables 1, 2, 3, respectively, but having assumed for 
all galaxies a Scalo IMF. 
In Tables 1-6 we provide also the comoving density of $^{4}He$ in the stars, ISM and IGM. 
Here we calculate only the amount of $^{4}He$ synthesized by stars, hence our values exclude the $^{4}He$ produced by 
Big Bang nucleosynthesis. In section 3.3.1 we will use our estimated abundances of newly produced $^{4}He$ 
to infer its primordial abundance $Y_{p}$.

\begin{table*}
\begin{flushleft}
\caption[]{Calculated comoving densities, expressed in $M_{\odot}/Mpc^{3}$,  for various elements for ellipticals, in the case of a Salpeter IMF. 
The elements produced by elliptical galaxies are locked into three phases: stars, ISM gas and IGM. 
Column 1: symbol of the chemical element; 
columns 2-5: mass densities in of the $i-$th chemical element in stars, ISM gas, IGM and the total comoving mass density.}
\begin{tabular}{l|l|l|l|l}
\noalign{\smallskip}
\hline
\hline
\noalign{\smallskip}
element & $\rho_{i, \, E, \, *}$ & $\rho_{i, \, E, \, ISM }$ & $\rho_{i, \, E, \, IGM}$ & $\rho_{i, \, E, \, Tot}$ \\
\noalign{\smallskip}
\hline
\noalign{\smallskip}
$^{4}$He & $6.60 \cdot 10^{5}$  & 1005  & $4.43 \cdot 10^{6}$ & $5.09 \cdot 10^{6}$  \\
C & $9.23 \cdot 10^{4}$ & 27 & $5.47\cdot 10^{5}$ & $6.39\cdot 10^{5}$  \\ 
O &$1.21\cdot 10^{6}$  & 169 & $1.87\cdot 10^{ 6}$  & $3.08\cdot 10^{6}$  \\
N &$7.91\cdot 10^{4}$ & 17 & $2.11\cdot 10^{5}$ & $2.90\cdot 10^{5}$  \\ 
Fe & $6.22\cdot 10^{4}$ & 50 & $3.35\cdot 10^{5}$ & $3.97\cdot 10^{5}$ \\ 
Si & $6.60\cdot 10^{4}$ & 24 & $1.52\cdot 10^{5}$ & $2.18\cdot 10^{5}$ \\ 
Mg & $5.27\cdot 10^{4}$& 12 & $8.6\cdot 10^{4}$& $1.39\cdot 10^{5}$ \\ 
Zn & 145 & 0.07 & 479 & 625 \\
Z & $1.87\cdot 10^{6}$& 402 & $3.61\cdot 10^{6}$& $5.48\cdot 10^{6}$  \\ 
\hline
\hline
\end{tabular}
\end{flushleft}
\end{table*}
\begin{table*}
\begin{flushleft}
\caption[]{Calculated comoving densities, expressed in $M_{\odot}/Mpc^{3}$, for various elements for spirals, in the case of a Salpeter IMF. 
The elements produced by spiral galaxies are locked into two phases: stars and ISM gas. 
It is assumed that spirals do not expel any fraction of their mass into the IGM. 
Column 1: symbol of the chemical element; 
columns 2-4: mass densities in of the $i-$th chemical element in stars, ISM gas and the total comoving mass density.}
\begin{tabular}{l|l|l|l|l}
\noalign{\smallskip}
\hline
\hline
\noalign{\smallskip}
element & $\rho_{i, \, S, \, *}$ & $\rho_{i, \, S, \, ISM }$ & $\rho_{i, \, S, \, Tot}$ \\
\noalign{\smallskip}
\hline
\noalign{\smallskip}
$^{4}$He & $1.77 \cdot 10^{6}$ &  $7.28 \cdot 10^{5}$& $2.50 \cdot 10^{6}$   \\
C & $2.59 \cdot 10^{5}$& $8.16 \cdot 10^{4}$& $3.41 \cdot 10^{5}$ \\
O & $1.93\cdot 10^{6}$ & $3.70 \cdot 10^{5}$& $2.30\cdot 10^{6}$\\
N & $1.41\cdot 10^{5}$& $3.64 \cdot 10^{4}$ & $1.77\cdot 10^{5}$ \\
Fe & $1.93\cdot 10^{5}$& $4.66\cdot 10^{4}$ & $2.40\cdot 10^{5}$\\
Si & $1.30\cdot 10^{5}$& $2.70 \cdot 10^{4}$& $1.57\cdot 10^{5}$\\
Mg & $8.96\cdot 10^{4}$ & $1.75 \cdot 10^{4}$& $1.07\cdot 10^{5}$ \\
Zn & 424 & 138 & 562 \\
Z & $3.05\cdot 10^{6}$& $6.56\cdot 10^{5}$&  $3.71 \cdot 10^{6}$ \\ 
\hline
\hline
\end{tabular}
\end{flushleft}
\end{table*}
\begin{table*}
\begin{flushleft}
\caption[]{Calculated comoving densities, expressed in $M_{\odot}/Mpc^{3}$, for various elements for irregulars, in the case of a Salpeter IMF. 
The elements produced by irregular galaxies are locked into two phases: stars and ISM gas. 
It is assumed that irregulars do not expel any fraction of their mass into the IGM. 
Column 1: symbol of the chemical element; 
columns 2-4: mass densities in of the $i-$th chemical element in stars, ISM gas and the total comoving mass density.}
\begin{tabular}{l|l|l|l|l}
\noalign{\smallskip}
\hline
\hline
\noalign{\smallskip}
element & $\rho_{i, \, I, \, *}$ & $\rho_{i, \, I, \, ISM }$ & $\rho_{i, \, I, \, Tot}$ \\
\noalign{\smallskip}
\hline
\noalign{\smallskip}
$^{4}$He & $4.86 \cdot 10^{3}$  & $1.08 \cdot 10^{5}$  & $1.13 \cdot 10^{5}$ \\
C & 971 & $1.60 \cdot 10^{4}$ & $1.70 \cdot 10^{4}$\\
O & $6.12\cdot 10^{3}$ & $9.74 \cdot 10^{4}$& $1.03 \cdot 10^{5}$\\
N & 388 & $7.03 \cdot 10^{3}$  & $7.42 \cdot 10^{3}$ \\
Fe & 631 & $1.08 \cdot 10^{4}$ & $1.14 \cdot 10^{4}$\\
Si & 389 & $5.95 \cdot 10^{3}$ & $6.34 \cdot 10^{3}$ \\
Mg & 243 & $4.33 \cdot 10^{3}$ & $4.57 \cdot 10^{3}$ \\
Zn & 0.78 & 13.52 & 14.3 \\
Z & $9.71\cdot 10^{3}$& $1.60 \cdot 10^{5}$& $1.70\cdot 10^{5}$ \\ 
\hline
\hline
\end{tabular}
\end{flushleft}
\end{table*}
\begin{table*}
\begin{flushleft}
\caption[]{Calculated comoving densities, expressed in $M_{\odot}/Mpc^{3}$, for various elements for ellipticals, 
in the case of a Scalo IMF. 
The elements produced by elliptical galaxies are locked into three phases: stars, ISM gas and IGM. 
Column 1: symbol of the chemical element; 
columns 2-5: mass densities in of the $i-$th chemical element in stars, ISM gas, IGM and the total comoving mass density.}
\begin{tabular}{l|l|l|l|l}
\noalign{\smallskip}
\hline
\hline
\noalign{\smallskip}
element & $\rho_{i, \, E, \, *}$ & $\rho_{i, \, E, \, ISM }$ & $\rho_{i, \, E, \, IGM}$ & $\rho_{i, \, E, \, Tot}$ \\
\noalign{\smallskip}
\hline
\noalign{\smallskip}
$^{4}$He & $4.93 \cdot 10^{5}$  & 870 &  $1.77 \cdot 10^{6}$  & $2.26 \cdot 10^{6}$ \\
C & $2.93\cdot 10^{4}$ & 12 & $3.2\cdot 10^{5}$ & $3.50\cdot 10^{5}$  \\
O & $3.38\cdot 10^{5}$ & 80 & $4.58\cdot 10^{5}$ & $7.96\cdot 10^{5}$  \\
N & $3.83\cdot 10^{4}$ & 9 &$8.27\cdot 10^{4}$ & $1.21\cdot 10^{5}$  \\
Fe &$2.16\cdot 10^{4}$  & 35 & $1.69\cdot 10^{5}$ & $1.91\cdot 10^{5}$  \\
Si & $2.08\cdot 10^{4}$ & 10 & $5.20\cdot 10^{4}$ & $7.28\cdot 10^{4}$  \\
Mg & $1.51\cdot 10^{4}$  & 4 & $2.03\cdot 10^{4}$ & $3.54\cdot 10^{4}$  \\
Zn & 44 & 0.03 & $1.77 \cdot 10^{2}$ & $2.21\cdot 10^{2}$  \\
Z  & $5.23\cdot 10^{5}$ & 169 & $1.18\cdot 10^{6}$ & $1.7\cdot 10^{6}$ \\
\hline
\hline
\end{tabular}
\end{flushleft}
\end{table*}
\begin{table*}
\begin{flushleft}
\caption[]{Calculated comoving densities, expressed in $M_{\odot}/Mpc^{3}$, for various elements for spirals, in the case of a Scalo IMF. 
The elements produced by spiral galaxies are locked into two phases: stars and ISM gas. 
It is assumed that spirals do not expel any fraction of their mass into the IGM. 
Column 1: symbol of the chemical element; 
columns 2-4: mass densities in of the $i-$th chemical element in stars, ISM gas and the total comoving mass density.}
\begin{tabular}{l|l|l|l|l}
\noalign{\smallskip}
\hline
\hline
\noalign{\smallskip}
element & $\rho_{i, \, S, \, *}$ & $\rho_{i, \, S, \, ISM }$ & $\rho_{i, \, S, \, Tot}$ \\
\noalign{\smallskip}
\hline
\noalign{\smallskip}
$^{4}$He & $1.12 \cdot 10^{6}$  & $5.36 \cdot 10^{5}$ & $1.65 \cdot 10^{6}$ \\
C & $1.86 \cdot 10^{5}$ & $6.57  \cdot 10^{4}$ & $2.52 \cdot 10^{5}$\\
O & $7.03 \cdot 10^{5}$& $1.34 \cdot 10^{5}$& $8.37 \cdot 10^{5}$\\
N & $8.25 \cdot 10^{4}$ & $1.91  \cdot 10^{4}$ &  $1.02 \cdot 10^{5}$\\
Fe & $1.15  \cdot 10^{5}$ & $2.89  \cdot 10^{4}$ & $ 1.44 \cdot 10^{5}$\\
Si & $5.70  \cdot 10^{4}$ &  $1.19 \cdot 10^{4}$ &  $6.89 \cdot 10^{4}$\\
Mg & $3.29 \cdot 10^{4}$ & $6.11 \cdot 10^{3}$ & $3.90 \cdot 10^{4}$\\
Zn & 198 & 61 & 259  \\
Z &  $1.24 \cdot 10^{6}$ & $2.95 \cdot 10 ^{5}$ & $1.54 \cdot 10 ^{6}$\\ 
\hline
\hline
\end{tabular}
\end{flushleft}
\end{table*}
\begin{table*}
\begin{flushleft}
\caption[]{Calculated comoving densities, expressed in $M_{\odot}/Mpc^{3}$, for various elements for irregulars, in the case of a Scalo IMF. 
The elements produced by irregular galaxies are locked into two phases: stars and ISM gas. 
It is assumed that irregulars do not expel any fraction of their mass into the IGM. 
Column 1: symbol of the chemical element; 
columns 2-4: mass densities in of the $i-$th chemical element in stars, ISM gas and the total comoving mass density.}
\begin{tabular}{l|l|l|l|l}
\noalign{\smallskip}
\hline
\hline
\noalign{\smallskip}
element & $\rho_{i, \, I, \, *}$ & $\rho_{i, \, I, \, ISM }$ & $\rho_{i, \, I, \, Tot}$ \\
\noalign{\smallskip}
\hline
\noalign{\smallskip}
$^{4}$He & $3.40 \cdot 10^{3}$ & $7.47 \cdot 10^{4}$  & $7.81 \cdot 10^{4}$  \\
C & 826 & $1.43 \cdot 10^{4}$  & $1.51 \cdot 10^{4}$ \\
O & $2.51 \cdot 10^{3}$ & $4.08 \cdot 10^{4}$ & $4.33 \cdot 10^{4}$\\
N & 342 & $5.56 \cdot 10^{3}$& $5.90 \cdot 10^{3}$\\
Fe & 461 & $8.04 \cdot 10^{3}$ & $8.50 \cdot 10^{3}$\\
Si & 203 & $3.38 \cdot 10^{3}$ & $3.58 \cdot 10^{3}$\\
Mg & 109 & $1.77 \cdot 10^{3}$ & $1.87 \cdot 10^{3}$\\
Zn & 0.5 & 8.7 & 9.2 \\
Z & $4.85 \cdot 10^{3}$ & $8.03 \cdot 10^{4}$ & $8.52 \cdot 10^{4}$\\
\hline
\hline
\end{tabular}
\end{flushleft}
\end{table*}

\subsubsection{Mass density of the heavy elements in spheroids, discs and irregulars}
Tables 1-6 indicate that, in the case of a Salpeter (Scalo) IMF, spheroids expel the $66 \%$ ($69 \%$) of their metals into the IGM, whereas 
practically all the remainder is locked up into stars and remnants. Spiral galaxies incorporate into stars $\sim 82 \%$ ($\sim 80 \%$) 
of their metals. For both IMFs, the big majority (94 $\%$) of the metals in irregulars lies in the ISM. Our results indicate that 
the assumption of a universal Salpeter IMF produces  
a total metal mass density $\rho_{Z}=9.37 \cdot 10^{6} M_{\odot}/Mpc^{3}$, whereas with a Scalo IMF we obtain 
$\rho_{Z}=3.32 \cdot 10^{6} M_{\odot}/Mpc^{3}$, i.e. roughly $1/3$ of the value obtained with the Salpeter IMF. 
The principal metal producers in the universe are the spheroids where, 
for a Salpeter IMF, $58 \%$ of the total cosmic amount of metals is built. 
Spiral galaxies also represent important metal producers, contributing to the $40 \%$ of the total metal budget, 
whereas irregulars represent negligible contributors, only with the $2 \%$. 
Of the total comoving density of metals, $60 \%$ is represented by oxygen, which is the most abundant element 
in the universe after hydrogen and $^{4}$He. Fe is often considered a reliable metallicity tracer and is 
widely measured in clusters of galaxies (Renzini et al. 1993, Renzini 2003) and in Damped Lyman-$\alpha$ (DLA) 
systems (Lu et al. 1996, Prochaska \& Wolfe 2002), though suffering substantial depletion into dust grains 
(Vladilo 1998). According to our results, it contributes to only the $7 \%$ of the total metal comoving density. 
Another element often used as metallicity tracer in DLAs is zinc (Pettini et al. 1997, Pettini et al. 1999, Vladilo et al. 2000).  
Our calculation indicate that its contribution is very small, namely the $0.1 \%$ of the total. 
\begin{table*}
\begin{flushleft}
\caption[]{Predicted value of $\Omega_{Z}$ (i.e. the total mass density of all the metals divided by the critical 
density of the universe) at $z=0$ for a Salpeter and a Scalo IMF and compared with previous 
determinations by various authors.}
\begin{tabular}{l|l|l}
\noalign{\smallskip}
\hline
\hline
\noalign{\smallskip}
Author & $\Omega_{Z}$ & $\rho_{Z} (M_{\odot}/Mpc^{3})$\\
\hline
\noalign{\smallskip}
Madau et al. 1996 & $7.8\cdot 10^{-5}$ &   $5.4 \cdot 10^{6}$ \\
Zepf \& Silk 1996 & $5.8 \cdot 10^{-4}$ & $4. \cdot 10^{7}$ \\
Mushotzky \& Loewenstein 1997 &  $2. \cdot 10^{-4}$ & $1.4 \cdot 10^{7}$   \\
Dunne et al. 2003 & $ 9.6 \cdot 10^{-5} - 1.9 \cdot 10^{-4}$ & $6.7 - 13.3  \cdot 10^{6}$  \\
Pagel 2002    &  $2.5 \cdot 10^{-4}$ &  $1.7  \cdot 10^{7}$ \\
Finoguenov et al. 2003 &  $ 9.9 \cdot 10^{-5} - 2.00 \cdot 10^{-4}$  & $ 6.9 - 13.8 \cdot 10^{6}$ \\
present work (Salpeter IMF) & $ 1.35 \cdot 10^{-4}$ & $9.37 \cdot 10^{6}$\\
present work (Scalo IMF) & $4.8 \cdot 10^{-5}$ & $3.32 \cdot 10^{6}$\\
\noalign{\smallskip}
\hline
\hline
\end{tabular}
\end{flushleft}
\end{table*}
In Table 7 we show our predicted values of $\Omega_{Z}$, namely the total metal comoving density calculated at the present 
time divided by the 
critical density of the universe ($\rho_{c}=6.94 \cdot 10 ^{10} M_{\odot}/Mpc^{3}$ for the value of $h=0.5$ adopted here), 
along with other estimates performed previously by various authors. 
In the case of a Salpeter IMF, we predict a total metal comoving density of   
$\rho_{Z}=9.37 \cdot 10^{6} M_{\odot}/Mpc^{3}$, higher than the value of $5.4 \cdot 10^{6} M_{\odot}/Mpc^{3}$ computed by Madau et al. (1996). 
A possible reason for this discrepancy could be the fact that, as argued by Mushotzky \& Loewenstein (1997), the galaxy sample collected by Madau et al. 
is biased towards blue star forming disc galaxies in the field and excludes the protospheroids forming stars in obscuring 
environments, 
with a consequent underestimation of the total SFR and metal production.\\
However, our value is smaller than $\rho_{Z}=1.4 \cdot 10^{7} M_{\odot}/Mpc^{3}$ inferred by Mushotzky \& Loewenstein (1997) on the 
basis of observations of clusters of galaxies and than $\rho_{Z}= 4 \cdot 10^{7} M_{\odot}/Mpc^{3}$ calculated by Zepf \& Silk (1996) for 
cluster ellipticals. 
This discrepancy is not due to the luminosity function  
parameters chosen for the normalization of the galaxy population, being the ones by Loveday et al. (1992), adopted both by 
Mushotzky \& Loewenstein (1997) and Zepf \& Silk (1996), 
quite similar to the one adopted here, i.e. that of Marzke et al. (1998). In our opinion, 
the difference is more due to the fact that clusters represent biased (and not average) sites of metal production 
in the universe, for several  reasons. 
Firstly, they are known to grow in zones of strong overdensity, generally characterized by an enhancement of star formation 
and metal production with respect to the field (Cen \& Ostriker 1999). 
Secondly, the galaxy proportions in clusters are unrepresentative 
of the average galaxy population, since in clusters a very high percentage of galaxies is represented by spheroids 
(up to $80 \%$, Renzini et al. 1993), whereas studies based on the optical luminosity function 
indicate that $\sim 50 \%$ of the field galaxy population have discs (FHP98).\\ 
Furthermore, from a detailed analysis of the budget of baryons at $z \sim 0$, we know that only $ 12.5 \%$ 
of the galaxies are in clusters, 
whereas the vast majority is in the field (FHP98). For all these reasons, cluster galaxies cannot be 
considered as representative of the whole galaxy population, rather they represent fair samples of the universe only 
by the chemical evolution point of view, since they have retained the totality of the baryons out of which they formed, 
hence they are the best sites where the closed-box approximation can hold (Renzini 1997).\\
Another reason for the discrepancy between our value and the one calculated by Mushotzky \& Loewenstein could 
be due to the choice of the IMF: this aspect and the effects of the adoption of a flatter IMF are described in section 1.3.\\
The reason for the stronger discrepancy between our value and the one by Zepf \& Silk (1996) resides instead both 
in the shape of the IMF 
and in the adopted stellar yields: 
those authors have adopted a Salpeter IMF with a mass cut at $m=3 M_{\odot}$ and the yields by Woosley \& Weaver (1995) 
for massive stars, and Renzini \& Voli (1981) for low and 
intermediate mass stars. According to their results, a total mass in metals of $0.42$ 
per solar mass of remnant is produced, whereas in our case the ``overall true  yield'', defined as the mass of newly 
synthesized and ejected metals divided by the mass in long-living stars and remnants (Pagel 2001) is $\sim 0.036$.  
Moreover, Pagel (2002) considered metal production in clusters and  calculated an overall true yield of $2.6 Z_{\odot}=0.05$.\\
Our $\Omega_{Z}$ value calculated with the Salpeter IMF is instead in good agreement with the ones provided by Dunne et al. (2003) and 
Finoguenov et al. (2003).\\
With the assumption of a universal Scalo-like IMF, the total amounts of metals 
in the local universe is $\rho_{Z}=3.32 \cdot 10^{6} M_{\odot}/Mpc^{3}$, a value which is in disagreement with any other estimate. 
On the other hand, if we adopt an IMF varying among the different morphological types, 
namely a combination of the Scalo IMF for galactic discs and a Salpeter one for spheroids and irregulars, 
we obtain a total present amount of metals of $ 7.2 \cdot 10^{6} M_{\odot}/Mpc^{3}$, close to the lower limits 
derived by Dunne et al. (2003) and Finoguenov et al. (2003). 
Our results therefore suggest that, \emph {if the IMF were universal, the Salpeter one allows us to reproduce 
the constraints on the total metal budget of the local universe, better than the Scalo one.} 
\begin{table*}
\begin{flushleft}
\caption[]{Calculated comoving densities, expressed in $M_{\odot}/Mpc^{3}$,  for various elements for ellipticals, in the case of an Arimoto Yoshii IMF. 
The elements produced by elliptical galaxies are locked into three phases: stars, ISM gas and IGM. 
Column 1: symbol of the chemical element; 
columns 2-5: mass densities in of the $i-$th chemical element in stars, ISM gas, IGM and total.}
%

\begin{tabular}{l|l|l|l|l}
\noalign{\smallskip}
\hline
\hline
\noalign{\smallskip}
element & $\rho_{i, \, E, \, *}$ & $\rho_{i, \, E, \, ISM }$ & $\rho_{i, \, E, \, IGM}$ & $\rho_{i, \, E, \, Tot}$ \\
\noalign{\smallskip}
\hline
\noalign{\smallskip}
$^{4}$He & $5.03 \cdot 10^{6}$  & $2.4 \cdot 10^{3}$ & $2.11 \cdot 10^{7}$ &  $2.62 \cdot 10^{7}$ \\
C & $2.65 \cdot 10^{5}$ &  77 & $ 2.11 \cdot 10^{6}$  & $ 2.38 \cdot 10^{6}$\\ 
O &$3.93 \cdot 10^{6}$  & 880  & $1.93 \cdot 10^{7}$  &  $2.32 \cdot 10^{7}$ \\
N &$2.97 \cdot 10^{5}$ &  66 & $2.39 \cdot 10^{6}$  & $ 2.69 \cdot 10^{6}$ \\
Fe & $1.23 \cdot 10^{5}$ & 131  & $1.32 \cdot 10^{6}$ &  $1.44 \cdot 10^{6}$\\
Si & $2.06 \cdot 10^{5}$ & 66 & $1.19 \cdot 10^{6}$ & $1.40 \cdot 10^{6}$ \\ 
Mg & $1.80 \cdot 10^{5}$& 44 & $9.3 \cdot 10^{5}$  &  $1.11 \cdot 10^{6}$ \\ 
Zn & 667 & 0.22 & 5784  & $6450.88$ \\
Z & $5.67 \cdot 10^{6}$& $1.51 \cdot 10^{3}$  & $3.13 \cdot 10^{7}$   &  $3.70 \cdot 10^{7}$ \\ 
\hline
\hline
\end{tabular}
\end{flushleft}
\end{table*}

%
%

\subsubsection{Effects of a flat IMF in spheroids} 
Spheroids contain the oldest stellar populations in the universe, with ages up to $\sim 12-13$ Gyr for Galactic 
globular clusters and Bulge stars (Renzini 1994). The initial mass function governing the formation 
of such stars is currently a matter of debate.\\
A higher average temperature, and consequently a larger Jeans mass require that in the past the stars 
could have formed according to a mass distribution function different from the the present one, 
i.e. characterized by a flatter slope ($x < 1.35$). The effect is an overproduction of massive stars and 
a underproduction of solar-like mass stars, which represent the bulk of the stellar population in 
present-day spheroids. Some observational facts seem to point towards this evidence: 
for instance, to explain the Fe mass-to-light ratio observed in clusters of galaxies, 
Arnaud et al. (1992) suggested the existence of a bimodal IMF in cluster ellipticals, i.e. Salpeter-like 
for quiescent star formation and flatter for bursts, with the production of only massive stars. 
However, Gibson (1996) has demonstrated that such an IMF is disfavored by chemical and photometric  
arguments, such as the existence of subsolar metallicity components in nearby ellipticals, and by the 
fact that it cannot account for the color-magnitude diagram observed in cluster ellipticals.\\ 
The IMF proposed by Arimoto \& Yoshii (AY) (1987), characterized by a power-law with a $x=0.95$ slope, 
can account for some photometric features of cluster ellipticals, such as the U-B and V-R 
colors of giant ellipticals. We have recalculated the metal production rates for spheroids with such an IMF, and 
in Table 8  
we show our results for the comoving densities of various elements in stars, ISM and IGM. 
With an AY IMF, we find that $\sim 85 \%$ of the metals are ejected by spheroids into the IGM, whereas practically all 
the remainder is locked up into stars and remnants.  
The AY IMF leads to an excessively high metal content in 
spheroids at the present time, and to an overestimation of the metal comoving density, in excess of the value of  
Mushotzky \& Loewenstein by a factor of $2.6$. Other recent arguments suggest that such a flat IMF is inadequate for 
ellipticals in clusters, since it overproduces the abundance of Fe and the $[O/Fe]$ ratio observed 
in the Intra-cluster medium (Pipino et al. 2002). 
We conclude that a $x=0.95$ slope for the stellar IMF of spheroids represents an unlikely scenario,  
although an intermediate slope between the Salpeter and the AY cannot be excluded. 

\begin{table*}
\begin{flushleft}
\caption[]{Mean $^{4}$He, C, O, N, Fe, Si, Mg, Zn abundances (by mass) and total metallicity $Z$ 
for various galactic types in stars and ISM  for a Salpeter IMF. 
First, second and third  columns: average stellar 
abundances in ellipticals, spirals and irregulars, respectively. Fourth, fifth and sixth columns: 
average ISM abundances in ellipticals, spirals and irregulars, respectively.  }
\begin{tabular}{l|l|l||l|l|l}
\noalign{\smallskip}
\hline
\hline
\noalign{\smallskip}
{\tiny $<X_{He}>_{*,E}$} & {\tiny $<X_{He}>_{*,S}$} & {\tiny $<X_{He}>_{*,I}$} & {\tiny $<X_{He}>_{g,E}$} & 
{\tiny $<X_{He}>_{g,S}$} & {\tiny $<X_{He}>_{g,I}$} \\ 
\noalign{\smallskip}
\hline
\noalign{\smallskip}
 0.005  & 0.015   & 0.001 & 0.05  & 0.05 & 0.002 \\
\hline
\hline
\noalign{\smallskip}
{\tiny $<X_{C}>_{*,E}$} & {\tiny $<X_{C}>_{*,S}$} & {\tiny $<X_{C}>_{*,I}$} &{\tiny $<X_{C}>_{g,E}$} & {\tiny $<X_{C}>_{g,S}$} & {\tiny $<X_{C}>_{g,I}$} \\
\noalign{\smallskip}
\hline
\noalign{\smallskip}
 0.0007 & 0.0022 & 0.0002 &0.0013 & 0.0056 & 0.00029 \\
\hline
\hline
\noalign{\smallskip}
{\tiny $<X_{O}>_{*,E}$} & {\tiny $<X_{O}>_{*,S}$} & {\tiny $<X_{O}>_{*,I}$} & {\tiny $<X_{O}>_{g,E}$} & {\tiny $<X_{O}>_{g,S}$} & {\tiny $<X_{O}>_{g,I}$} \\
\noalign{\smallskip}
\hline
\noalign{\smallskip}
 0.0092 & 0.0164 & 0.00126 & 0.0084 & 0.0254 & 0.0018 \\
\hline
\hline
{\tiny $<X_{N}>_{*,E}$} & {\tiny $<X_{N}>_{*,S}$} & {\tiny $<X_{N}>_{*,I}$} & {\tiny $<X_{N}>_{g,E}$} & {\tiny $<X_{N}>_{g,S}$} & {\tiny $<X_{N}>_{g,I}$} \\
\noalign{\smallskip}
\hline
\noalign{\smallskip}
 0.0006 & 0.0012 & 0.00008 & 0.0008  & 0.0025 & 0.00013 \\
\hline
\hline
\noalign{\smallskip}
{\tiny $<X_{Fe}>_{*,E}$ } & {\tiny $<X_{Fe}>_{*,S}$ } & {\tiny $<X_{Fe}>_{*,I}$ } & {\tiny $<X_{Fe}>_{g,E}$ } & {\tiny $<X_{Fe}>_{g,S}$ } & {\tiny $<X_{Fe}>_{g,I}$ } \\
\noalign{\smallskip}
\hline
\noalign{\smallskip}
 0.000472 & 0.00164 & 0.00013 & 0.00242 & 0.0032 & 0.0002 \\
\hline
\hline
{\tiny $<X_{Si}>_{*,E}$} & {\tiny $<X_{Si}>_{*,S}$} & {\tiny $<X_{Si}>_{*,I}$} & {\tiny $<X_{Si}>_{g,E}$} & {\tiny $<X_{Si}>_{g,S}$} & {\tiny $<X_{Si}>_{g,I}$} \\
\noalign{\smallskip}
\hline
\noalign{\smallskip}
 0.0005  & 0.0011 & 0.00008 & 0.0012  & 0.00185 & 0.00011  \\
\hline
\hline
{\tiny $<X_{Mg}>_{*,E}$} & {\tiny $<X_{Mg}>_{*,S}$} & {\tiny $<X_{Mg}>_{*,I}$} & {\tiny $<X_{Mg}>_{g,E}$} & {\tiny $<X_{Mg}>_{g,S}$} & {\tiny $<X_{Mg}>_{g,I}$} \\
\noalign{\smallskip}
\hline
\noalign{\smallskip}
 0.0004  & 0.00076 & 0.00005 & 0.0006  &  0.0012 & 0.00008 \\
\hline
\hline
{\tiny $<X_{Zn}>_{*,E}$} & {\tiny $<X_{Zn}>_{*,S}$} & {\tiny $<X_{Zn}>_{*,I}$} & {\tiny $<X_{Zn}>_{g,E}$} & {\tiny $<X_{Zn}>_{g,S}$} & {\tiny $<X_{Zn}>_{g,I}$} \\
\noalign{\smallskip}
\hline
\noalign{\smallskip}
 $1.1 \cdot 10^{-6}$ & $3.6 \cdot 10^{-6}$  & $1.6 \cdot 10^{-7}$ &  $3.2 \cdot 10^{-6}$ &  $9.5 \cdot 10^{-6}$& $2.5 \cdot 10^{-7}$  \\
\hline
\hline
\noalign{\smallskip}
{\tiny $<Z>_{*,E}$ }& {\tiny $<Z>_{*,S}$ } & {\tiny $<Z>_{*,I}$ } & {\tiny $<Z>_{g,E}$ } & {\tiny $<Z>_{g,S}$ } & {\tiny $<Z>_{g,I}$ }  \\
\noalign{\smallskip}
\hline
\noalign{\smallskip}
 0.014 & 0.026 & 0.002 & 0.02 & 0.045 & 0.003 \\
\noalign{\smallskip}
\hline
\hline
\end{tabular}
\end{flushleft}
\end{table*}
\begin{table*}
\begin{flushleft}
\caption[]{Mean $^{4}$He, C, O, N, Fe, Si, Mg, Zn abundances (by mass) and total metallicity $Z$ for various galactic types in stars and ISM  
for a Scalo IMF.
First, second and third  columns: average stellar
abundances in ellipticals, spirals and irregulars, respectively. Fourth, fifth and sixth columns: 
average ISM abundances in ellipticals, spirals and irregulars, respectively. }
\begin{tabular}{l|l|l||l|l|l}
\hline
\hline
\noalign{\smallskip}
{\tiny $<X_{He}>_{*,E}$} & {\tiny $<X_{He}>_{*,S}$} & {\tiny $<X_{He}>_{*,I}$} & {\tiny $<X_{He}>_{g,E}$} & 
{\tiny $<X_{He}>_{g,S}$} & {\tiny $<X_{He}>_{g,I}$} \\ 
\noalign{\smallskip}
\hline
\noalign{\smallskip}
 0.005  & 0.009  & 0.0006 & 0.060  & 0.04 & 0.0013 \\
\hline
\hline
\noalign{\smallskip}
{\tiny $<X_{C}>_{*,E}$} & {\tiny $<X_{C}>_{*,S}$} & {\tiny $<X_{C}>_{*,I}$} & {\tiny $<X_{C}>_{g,E}$} & {\tiny $<X_{C}>_{g,S}$} & {\tiny $<X_{C}>_{g,I}$} \\
\noalign{\smallskip}
\hline
\noalign{\smallskip}
 $0.0003$  & $0.0015$ & $0.00015$ & $0.0008$ & $0.0049$ & $0.00025$ \\
\hline
\hline
\noalign{\smallskip}
{\tiny $<X_{O}>_{*,E}$} & {\tiny $<X_{O}>_{*,S}$} & {\tiny $<X_{O}>_{*,I}$} & {\tiny $<X_{O}>_{g,E}$} & {\tiny $<X_{O}>_{g,S}$} & {\tiny $<X_{O}>_{g,I}$} \\
\noalign{\smallskip}
\hline
\noalign{\smallskip}
 $0.0034$ & $0.0057$ & $0.00044$ & $0.0055$ & $0.01$ & $0.00071$ \\
\hline
\hline
{\tiny $<X_{N}>_{*,E}$} & {\tiny $<X_{N}>_{*,S}$} & {\tiny $<X_{N}>_{*,I}$} & {\tiny $<X_{N}>_{g,E}$} & {\tiny $<X_{N}>_{g,S}$} & {\tiny $<X_{N}>_{g,I}$} \\
\noalign{\smallskip}
\hline
\noalign{\smallskip}
 $0.0004$ & $0.00067$ & $0.000061$ & $0.00062$  & $0.0014$ & $0.000097$ \\
\hline
\hline
\noalign{\smallskip}
{\tiny $<X_{Fe}>_{*,E}$ } & {\tiny $<X_{Fe}>_{*,S}$ } & {\tiny $<X_{Fe}>_{*,I}$ } & {\tiny $<X_{Fe}>_{g,E}$ } & {\tiny $<X_{Fe}>_{g,S}$ } & {\tiny $<X_{Fe}>_{g,I}$ } \\
\noalign{\smallskip}
\hline
\noalign{\smallskip}
 $0.00022$ & $0.0093$ & $0.00081$ & $0.00241$ & $0.00216$ & $0.00014$ \\
\hline
\hline
{\tiny $<X_{Si}>_{*,E}$} & {\tiny $<X_{Si}>_{*,S}$} & {\tiny $<X_{Si}>_{*,I}$} & {\tiny $<X_{Si}>_{g,E}$} & {\tiny $<X_{Si}>_{g,S}$} & {\tiny $<X_{Si}>_{g,I}$} \\
\noalign{\smallskip}
\hline
\noalign{\smallskip}
 $0.00021$  & $0.00046$ & $0.000036$ & $0.00071$  & $0.00089$ & $0.000059$ \\
\hline
\hline
{\tiny $<X_{Mg}>_{*,E}$} & {\tiny $<X_{Mg}>_{*,S}$} & {\tiny $<X_{Mg}>_{*,I}$} & {\tiny $<X_{Mg}>_{g,E}$} & {\tiny $<X_{Mg}>_{g,S}$} & {\tiny $<X_{Mg}>_{g,I}$} \\
\noalign{\smallskip}
\hline
\noalign{\smallskip}
 $0.00015$  & $0.000265$ & $0.000019$ & $0.00026$  &  $0.00046$ & $0.000031$ \\
\hline
\hline
{\tiny $<X_{Zn}>_{*,E}$} & {\tiny $<X_{Zn}>_{*,S}$} & {\tiny $<X_{Zn}>_{*,I}$} & {\tiny $<X_{Zn}>_{g,E}$} & {\tiny $<X_{Zn}>_{g,S}$} & {\tiny $<X_{Zn}>_{g,I}$} \\
\noalign{\smallskip}
\hline
\noalign{\smallskip}
 $4.5 \cdot 10^{-7}$ & $1.6 \cdot 10^{-6}$  & $8.9 \cdot 10^{-8}$ & $2.3 \cdot 10^{-6}$ &  $4.5 \cdot 10^{-6}$ &  $1.5 \cdot 10^{-7}$ \\
\hline
\hline
\noalign{\smallskip}
{\tiny $<Z>_{*,E}$ }& {\tiny $<Z>_{*,S}$ } & {\tiny $<Z>_{*,I}$ } & {\tiny $<Z>_{g,E}$ } & {\tiny $<Z>_{g,S}$ } & {\tiny $<Z>_{g,I}$ } \\
\noalign{\smallskip}
\hline
\noalign{\smallskip}
 $0.0053$ & $0.01$ & $0.00086$ & $0.012$ & $0.022$ & $0.0014$ \\
\noalign{\smallskip}
\hline
\hline
\end{tabular}
\end{flushleft}
\end{table*}

\subsection{Average chemical abundances in stars and gas and the global metallicity of the universe} 
%
In Table 9 we show the mean metal abundances (by mass) in stars and in the ISM calculated for each galaxy type at the present time assuming a Salpeter IMF. 
In Table 10, we show the same quantities as in Table 9, but having assumed a Scalo IMF for all galaxies. 
The solar abundances adopted for reference are taken from Anders \& Grevesse (1989). 
In spheroids, for a Salpeter IMF the average stellar O abundance is practically solar, $<X_{O}>_{*,E}= 0.96 X_{O,\odot}$, 
the total average metallicity is 
$<Z>_{*,E}=0.74 Z_{\odot}$, whereas the average Fe abundance is $<X_{Fe}>_{*,E}=0.37 X_{Fe,\odot}$, 
in agreement with recent 
observations in nearby ellipticals by Kobayashi \& Arimoto (1999). 
These observations suggest values  spanning from $\sim 0.16$ to  $\sim 2$ times solar with 
an average value of $\sim 0.5$ solar. 
The predicted mean (O/Fe) ratio for stars in spheroids is $\sim 19.6$, namely 
 $[O/Fe] \sim 0.4$ dex, in very good agreement with the $[\alpha/Fe]$ values commonly derived for 
the dominant stellar populations of ellipticals (e. g., Thomas, Maraston \& Bender, 2002). 
On the other hand, our predicted (O/Fe) ratio in the gas of ellipticals is $3.47$, corresponding to 
$[O/Fe]\sim -0.33$ dex, consistent with recent Chandra and XMM-Newton data 
(e.g. Mushotsky 2002, Gastaldello \& Molendi 2002).\\
The average $[O/Fe]$ ratio in spirals and irregulars is $\sim 0.1$ dex for the stellar component and almost solar 
($\sim 0.01$ dex) for the gas component.\\
We note that our calculated mean abundances for spirals are higher than the ones inferred 
by Edmunds \& Phillips (1997), who find approximately solar values for both stars and gas. 
For the stellar abundances, we believe that this is partly due to the fact that their estimates are based 
on the assumption of an empirical law between metallicity and luminosity, i.e. $Z \propto \Sigma_{*}^{x}$, 
where $\Sigma_{*}$ is the surface-mass density of stars. 
Our metallicities represent mass-weighted (and not luminosity weighted ) averages, and 
chemical evolution models (Arimoto \& Yoshii 1987, Yoshii \& Arimoto 1987, Matteucci \& Tornamb\'e 1987) have shown that 
luminosity weighted metallicities are in general lower than mass weighted ones, since metal poor
giants tend to predominate in the visual luminosity, especially in low-mass spheroids.\\ 
On the other hand, for high-mass spheroids the two averages are almost coincident (Matteucci et al. 1998). 
Concerning the gas abundances in spirals, by means of our multi-zone model we have calculated mass-weighted averages on the whole 
disc, according to:\\
\begin{equation} 
<Z>_{g,S}=\frac {\sum_{j} \sigma_{j} Z_{j}}{\sum_{j} \sigma_{j}}\\
\end{equation} 
where the $\sigma_{j}$ are the total mass surface densities in the various regions of the disc and the $Z_{j}$ 
are the metallicities. Given our inside-out scenario for disc formation, the innermost regions 
are the ones experiencing the highest levels of metal enrichment and having the highest mass densities, 
hence the highest weights in the calculated average. 
Finally, we note that irregular systems have always undersolar abundances, both in the gas and in the stars. \\ 
As we can see from Table 10, the stellar and ISM abundances predicted for all galaxies having assumed a Scalo IMF are 
all smaller than in the case of the Salpeter IMF. In general, abundances for elliptical galaxies appear too 
small with respect to observations. For spirals, abundances calculated with the Scalo IMF (Table 10) are probably more consistent to 
 observational results (e.g., Vila-Costas \& Edmunds 1992) than in the case of the Salpeter IMF.\\ 
We can now calculate the mass-weighted mean metallicity of galaxies which, for a Salpeter IMF, is :\\
\begin{equation} 
<Z_{C}>=\frac{\rho_{g,C} \, Z_{g,C} + \rho_{*,C} \, Z_{*,C}}{\rho_{g,C} +\rho_{*,C} }= 0.0175  \\
\end{equation} 
We have then reached the result that the mean metallicity of 
\emph {galactic matter (i.e. stars and interstellar gas)} in the universe  
is $0.0175$, very close to the solar value. Therefore, in spite of the slight discrepancies 
due to the mass/luminosity averages, we confirm the result found by Edmunds \& Phillips (1997).\\

\subsubsection{Cosmic abundance of $^4$He: a hint on the primordial value $Y_{p}$ }

In Tables 9 and 10 we report only the stellar and ISM abundances in various galaxy types 
of the \underline{$^4$He newly produced in stars}, i.e. without the amount due to primordial nucleosynthesis. 
Our calculated abundances can then allow us a rough indirect derivation of the primordial $^4$He abundance $Y_{p}$. 
The astrophysical sites where the primordial $^4$He abundance can be derived are the extragalactic HII regions, 
whose average metallicities are in general highly undersolar ($Z < 0.1 Z_{\odot}$ or less, Pagel \& Edmunds 1981), 
since they have experienced the lowest levels of metal enrichment by stars. 
From the analysis of the total $^4$He in these systems and of the correlation between 
Y$_{obs}$ and the metallicity Z, in principle it is possible to derive Y$_{p}$ by performing 
an extrapolation of the Y$_{obs}$-Z relation at zero metallicity. 
Adopting this method, Olive et al. (1997) found $Y_{p}=0.234 \pm 0.002$, whereas Izotov et al. (1999) 
found $Y_{p}=0.2452 \pm 0.0015$. 
More recently, Peimbert et al. (2002) obtained a value of $Y_{p}=0.2384 \pm 0.0025$. 
An independent estimate of $Y_{p}$ is 
now provided by very recent results from WMAP (Spergel et al. 2003), suggesting a baryon to photon ratio 
 $\eta=6.1^{+0.3}_{-0.2} \cdot 10^{-10}$ which, assuming the Standard Big 
Bang Nucleosynthesis predictions by Cyburt, Fields \& Olive (2001), implies $Y_{p}=0.2484^{+0.0004}_{-0.0005}$. 
Pagel et al. (1992) have observed the $^4$He abundances in extragalactic HII regions of several local BCD galaxies. 
A straight mean of the measured $^{4}$He abundances yields a present time value of $Y=0.2456 \pm 0.012$. 
By subtracting to this value the $^{4}$He produced by irregular galaxies in the case of a Salpeter and a Scalo IMF,  
 we obtain $Y_{p}=0.2436 \pm 0.012$ and $Y_{p}=0.2443 \pm 0.012$, respectively. 
Given the uncertainty for the stellar IMF 
in local irregular galaxies, as fiducial value for the primordial $^{4}$He we assume the average value between the two, 
 $Y_{p}=0.244 \pm 0.012$. Such an estimate is in good agreement with all the appraisals of $Y_{p}$ based on the analysis of the 
helium content of extragalactic $HII$ regions discussed before, 
and also in agreement with $Y_{p} \simeq 0.243-0.244$ from Galactic globular 
clusters. We can also try to infer the primordial $^{4}$He abundance also by analyzing the galactic bulge. 
The stars in the bulge should have an average $^{4}$He abundance of $Y \sim 0.3-0.35$, as discussed by Renzini (1994). 
If we subtract to this value the $^{4}$He produced in spheroids, we obtain a rough estimate of $\sim 0.25-0.3$, which is higher than 
that obtained before but closer to the value derived by WMAP.

\renewcommand{\baselinestretch}{1.0}
\begin{table*}
\centering
\caption{Baryon and Metal Budget in the local universe, as predicted by our models and as suggested by 
the observations.}
\begin{tabular}{lccccc}
\\[-2.0ex] 
\hline
\\[-2.5ex]
\multicolumn{1}{l}{Component}&\multicolumn{2}{c}{$\Omega_{b}$}&\multicolumn{1}{c}{}&\multicolumn{2}{c}{$\Omega_{Z}$}\\
\multicolumn{1}{c}{}&\multicolumn{1}{c}{Observed}&\multicolumn{1}{c}{Predicted}&\multicolumn{1}{c}{}&\multicolumn{1}{c}{Observed}&\multicolumn{1}{c}{Predicted}\\
\hline
\hline
\\[-1.0ex]
stars in spheroids & $^a$$0.0036^{+0.00242}_{-0.0017}$  & 0.0019  &   & -  & $2.7\cdot 10^{-5}$ \\ 
stars in discs & $^a$$0.0012^{+0.0006}_{-0.00048}$   & 0.0017  &   & -  & $4.4\cdot 10^{-5}$ \\ 
stars in irregulars & $^a$$^a$$0.000096^{+0.000066}_{-0.000052}$ & 0.00007 &   & -  & $1.4\cdot 10^{-7}$ \\ 
\hline
Total stars &  $^a$$0.0049$  & $0.0037$  &   & $^{b,d,e}$$3.2-6.4 \cdot 10^{-5}$  & $7.1\cdot 10^{-5}$  \\ 
\hline
neutral gas in spheroids &   -  & $2.9\cdot 10^{-7}$  &   & -  &  $5.8\cdot 10^{-9}$ \\ 
neutral gas in discs &  $^f$$(2.5 \pm 0.8) \cdot 10^{-4}$  & $2.1\cdot 10^{-4}$  &   & -  & $9.45\cdot 10^{-6}$ \\ 
neutral gas in irregulars &  -  & $7.8\cdot 10^{-4}$  &   & -  & $2.3\cdot 10^{-6}$  \\ 
\hline
Total neutral Gas &  $^a$$0.0009 \pm 0.00012$  & $0.001$  &   & -  & $1.2\cdot 10^{-5}$ \\ 
\hline
IGM & $^a$$0.0239^{+0.0181}_{-0.01382}$  & $0.0126$  &   & $^c$$1.83 - 7.98 \cdot 10^{-5}$  & $5.2\cdot 10^{-5}$ \\ 
\hline
\hline
Total    & $^a$$0.02972^{+0.02867}_{-0.01393}$  & $0.0173$  &   &  $^c$ $1.15 - 2.02 \cdot 10^{-4}$   &  $ 1.35 \cdot 10^{-4}$ \\ 
\hline
\end{tabular}
\flushleft
References: $^a$ Fukugita, Hogan \& Peebles (1998), $^b$ Finoguenov et al. (2003), $^c$ Dunne et al. (2003), $^d$ Balogh et al. (2001), 
$^e$ Huang et al. (2003), $^f$ Salucci \& Persic (1999)\\
\end{table*}
\begin{table*}
\begin{flushleft}
\caption[]{Predicted IGM comoving densities (in units of the closure density of the universe) $\Omega_{i,IGM}$ and 
IGM average abundances by mass $X_{i,IGM}$ (in units of the solar abundances) for 
C, O, N, Fe, Si, Mg, Zn and for the global metallicity Z in the case of a Salpeter IMF. We assume the solar abundances by Anders \& Grevesse (1989), 
i.e. (by mass) $X_{C,\odot}=3.0 \cdot 10^{-3}$, 
$X_{O,\odot}=9.6 \cdot 10^{-3}$, $X_{N,\odot}=1.1 \cdot 10^{-3}$, $X_{Fe,\odot}=1.3 \cdot 10^{-3}$, $X_{Si,\odot}=7.1 \cdot 10^{-4}$, 
$X_{Mg,\odot}=5.1 \cdot 10^{-4}$, $X_{Zn,\odot}=2.1 \cdot 10^{-6}$, $Z=0.019$. }
\begin{tabular}{l|l|l}
\noalign{\smallskip}
\hline
\hline
\noalign{\smallskip}
Element & $\Omega_{i,IGM}$ & $X_{i,IGM}$\\
\hline
C & $7.9 \cdot 10^{-6}$  & 0.035 \\
O & $2.7 \cdot 10^{-5}$ & 0.04 \\
N & $3.0 \cdot 10^{-6}$& 0.04\\
Fe & $4.8 \cdot 10^{-6}$ & 0.05 \\
Si & $2.2 \cdot 10^{-6}$ & 0.04\\
Mg & $1.2 \cdot 10^{-6}$& 0.03\\
Zn & $6.9 \cdot 10^{-9}$ &  0.04\\
Z &  $5.2 \cdot 10^{-5}$& 0.04 \\
\noalign{\smallskip}
\noalign{\smallskip}
\hline
\hline

\end{tabular}
\end{flushleft}
\end{table*}

\subsubsection{Baryons and metals in the inter-galactic and intra-cluster medium}
In our evaluation of the global metallicity of the universe, we did not take into account the 
uncertain amount of primordial gas which constitutes the IGM and the intra-cluster medium (ICM)
\footnote{With the term Intra-cluster Medium we indicate the hot, ionized gas contained in clusters of galaxies, whereas    
with the designation inter-galactic medium, we mean the total amount of diffuse gas in the field. 
This latter is likely to represent 
a much higher fraction of the baryonic matter outside galaxies than the one represented by the intra-cluster gas.}.  
In general, intracluster plasma masses are well determined from X-ray observations (see FHP98 and references therein): 
within clusters of galaxies, the hot gas can represent a very high fraction of the total baryonic mass, 
in some cases up to $\sim 90\%$ (Matteucci \& Vettolani 1988, Matteucci \& Gibson 1995).  
The most popular metallicity tracer for the ICM is Fe: its X-ray lines are present in all clusters and groups, 
either warm or hot (Renzini 2003). Usually, the Fe abundances observed in the hot gas in clusters   
can span a relatively large interval: observations by various groups indicate a 
broad range of values from $\sim 0.2$ to $\sim 1 $  solar (Arnaud et al. 1992, Renzini 1997, 
Fukazawa et al. 1998, Finoguenov et al. 2001), with small or no-evolution up to $z \sim 0.5$ or more 
(Mushotzky \& Loewenstein 1997, Tozzi et al. 2003). 
On the other hand, the detected X-ray emission from inter-galactic plasma in galaxy groups and in the field is softer than in clusters, 
probably because of a lower virial temperature. This fact renders the observations of metals still more difficult than for the ICM. 
The field inter-galactic gas (IGM) dilutes the metals ejected by galaxies via winds and outflows, and is probably 
the astrophysical site presenting 
the lowest metal abundances ever observed, for example a Si abundance of $3.5 \times 10^{-4}$ of the solar value at $z \sim 5$ (Songaila 2001).\\
The mean Fe abundance of the IGM can be defined as:\\
\begin{equation} 
<X_{Fe,IGM}>=\frac{\Omega_{Fe,IGM}}{\Omega_{b,IGM}} \\
\end{equation} 
Therefore, by means of the computation of the Fe mass density expelled by galaxies and adopting a fiducial value 
for the mean Fe abundance of the IGM on the basis of current observations, in principle it is possible to 
provide an estimate of the baryonic density of the warm inter-galactic gas. 
We assume that all the Fe in the IGM has been ejected by spheroids through galactic winds 
(the contribution from winds in irregular galaxies is assumed to be negligible), and we 
evaluate the Fe density of the IGM by subtracting the Fe density in stars and in the ISM from 
the total amount of Fe produced up to now (see Table 1). In such a way, we obtain the value 
$\Omega_{Fe,IGM} = 4.8 \times 10^{-6}$. Being the total amount of Fe produced by spheroids 
$\Omega_{Fe, E} = 5.72 \times 10^{-6}$, this means that the $84 \%$ of all the Fe produced by 
spheroids (and exactly half of the total amount produced by all galaxies) is expelled into the IGM. 
This is due to the fact that in the galactic wind scenario, ellipticals develop an early wind (in our case 
a timescale between $0.3$ and 1 Gyr since their formation) emptying the galaxy of all the residual gas. 
At the occurrence of the wind, the bulk of the $\alpha$ elements 
 has already been produced and is locked up in stars. On the other hand, the production of the bulk of Fe has still 
to take place, 
since it occurs on longer timescales ($> 1 Gyr$) by means of type Ia SN explosions, so that all the 
Fe is restored into the ISM. Such explosions heat the inter-stellar gas so that is likely that 
minor galactic 
winds are triggered, with the consequent ejection of all the produced Fe into the IGM. 
For the mean Fe abundance of the IGM,  we assume a fiducial value of $0.3 X_{Fe, \odot}$. 
Such value has been inferred by Renzini (1997) assuming that the metal productivity of 
the stellar population is the same in the field as it is in galaxy clusters. 
This is equivalent to assume that the average ICM Fe abundance is representative of the whole IGM. 
This leads us to obtain a value of:\\
\begin{equation} 
\Omega_{b,IGM} \simeq  \frac{4.8 \times 10^{-6}}{0.00038} = 0.0126 \\ 
\end{equation} 
for the baryon density in the IGM. We stress that our calculated $\Omega_{b,IGM}$ 
 depends also on the luminosity function (LF) 
parameters and on the mean metallicity of the IGM, both factors suffering a certain degree of uncertainty. 
The uncertainty of the LF normalization, the most important parameter in our estimate, is about a factor of two (Cross et al. 2001), 
whereas the mean IGM Fe abundance is far more unconstrained. The possible overestimation of $<X_{Fe,IGM}>$ could lead us 
to severely underestimate the baryonic density in the IGM. Since the hot gas in the IGM represents the main 
contributor to the total baryon density, this has interesting consequences on the determination of $\Omega_{b}$, 
which will be discussed in the next section.

\subsubsection{Baryon and metal budget in the local universe} 
At this point, after having evaluated the mean metallicity of the universe in all its components, our aim 
is to assess the total amount of baryons in the local universe, along with the amounts of metals inside and 
outside galaxies. In Table 11, we present our results and we compare them with observations and 
other estimates by various authors. Some items (metal densities in stars in each galactic morphological type, 
neutral gas in spheroids and irregulars, metal densities in the ISM of spheroids, discs and irregulars) do not have 
observational values: for these components, the fields in Table 11 are left blank. 
First of all, we note that the agreement between our predictions and the observations for all the densities of the 
baryonic components is very good. We confirm the result that the vast majority of the baryons lies 
outside galaxies, yet at too low virial temperatures to be easily detected by current X-ray telescopes. 
Within galaxies, the majority of the baryons is in the form of stars. Our values for the stellar baryonic 
densities are in agreement within $1 \sigma $ with the values by FHP98, which indicate that the majority 
($\sim 73 \%$) of the stars at $z=0$ is in spheroids. However, our results suggest that the amount 
of stars in spheroids and disc galaxies is comparable, as several more recent estimates seem to indicate 
(Kochanek et al. 2001, Benson, Frenk and Sharples 2002). 
For the total baryon density, we obtain a value of $\Omega_{b}=0.0174$, again in agreement within $1 \sigma$ with the estimate by FHP98. 
However, measures of the abundance of D/H ratio in QSO absorbers provide a value of $\Omega_{b}=0.08$ 
(for $h=0.5$, O'Meara et al. 2001), and values from Cosmic Microwave Background data span from $0.036$ (Padin et al. 2001) 
up to the very high BOOMERANG value of $0.12$ (Lange et al. 2001). The recent WMAP results provide a value of 
$\Omega_{b}=0.0896 \pm 0.0009$ (Spergel et al. 2003). 
In Table 11 we report the value by 
FHP98 which is, as ours, largely based on the B luminosity function parameters 
(see also Persic \& Salucci 1992). 
The most important  parameter in such a determination, namely the normalization, is uncertain by a factor of two or more 
(Cross et al. 2001).  We believe that such an uncertainty can account for the discrepancy between our value 
and the ones present in the literature, along with the uncertainty in the IGM metallicity discussed above.
Assuming $\Omega_{b}=0.08$ as the fiducial value for the total baryonic density, and subtracting the total 
amount of baryons contained in galaxies, we obtain an IGM baryon density  
of $\Omega_{IGM}=0.0753$, and consequently a mean Fe abundance in the IGM  of $0.05$ solar.\\ 
Our results indicate that a value of $0.3$ solar, i.e. close to the values commonly observed in the ICM,    
for the mean Fe abundance of the IGM \emph {is too high to represent the true cosmic average}, 
i.e. that the average Fe abundance of the inter-galactic gas in the field should be lower than $0.3$ solar.\\
The average Fe IGM abundance is shown in Table 12, along with other predicted IGM metal abundances. All the metals studies here 
have similar abundances with respect to the solar ones, ranging from 0.03 (for Mg) to 0.05 (Fe) 
of the solar values. The predicted total IGM metallicity is $Z_{IGM}=0.04 Z_{\odot}$, whereas the average $[O/Fe]_{IGM}$ 
is $\sim -0.1 $ dex.\\
Most of the baryons lie outside galaxies, but this is not true for the majority of the metals: according to 
our predictions (and to uptodate evaluations, see Finoguenov et al. 2003, Dunne et al. 2003), the majority (the 
$\sim 52 \% $ in our case) is locked up in stars (the $\sim 51.9 \%$ in spheroids and spiral discs, the remaining $0.1 \%$ 
in small irregulars), whereas the $\sim 9 \%$ is in the ISM in galaxies. 
This means that $\sim 39 \%$ of metals has been ejected by galaxies into the IGM and ICM.
We stress that this is valid for the bulk of the metals (mainly $O$ and $\alpha-$elements), whereas Fe has a rather particular 
behaviour since it is produced on long timescales and $\sim 50 \%$ of the total amount of Fe is expelled by spheroids 
into the IGM/ICM. 
It is worth noting that our value of $\Omega_{Z,IGM}$ does not include the metals ejected into the IGM by dwarf galaxies, 
since in the present work we have assumed that all the metals present in the IGM have been expelled by large spheroids. 
In fact, Gibson \& Matteucci (1997) have demonstrated that dwarf galaxies cannot expel an amount 
of gas higher than the 15 \% of the 
total intra-cluster gas, 
whereas giant ellipticals are responsible for $\sim 20 \%$. 
This, together with the fact that dwarf galaxies produce (and expel) a negligible amount of 
metals (again Gibson \& Matteucci 1997), renders the former unsufficient contributors to the bulk of the IGM enrichment. 

\section{Conclusions}
By means of chemo-spectrophotometric models of galaxies of 
different morphological type we have carried out a detailed study of the history of 
element production in the universe. We divide the galaxy population into three categories: 
spheroids, discs and irregular systems. We normalize the galaxy fractions and densities according to the 
optical luminosity function in the local universe (Marzke et al. 1998), assuming that galaxies 
evolve in luminosity and not in number. Such a scenario is able to account for a large set of observables, 
such as the observed evolution of cosmic star formation and the galaxy luminosity density in various bands, 
along with the evolution of the cosmic supernova rate (CM03).\\
Our main results can be summarized as follows.\\
i) The assumption of a universal Scalo IMF implies  
a total metal mass density smaller ($\sim 1/3$) than the one obtained with a  Salpeter IMF. 
On the other hand, the total comoving metal density estimated assuming a Salpeter IMF in all galactic types 
is in good agreement with other estimates of the total amount of metals in the local universe, and 
suggests that the Salpeter IMF should be preferred over the Scalo one as a candidate for a possible universal IMF. 
However, a  Scalo-like IMF in discs, as indicated by observations in the Milky Way, and a Salpeter-like IMF 
in spheroids and irregulars, 
are still consistent with observational constraints. 
Our results indicate that the principal metal producers in the universe are the spheroids where, 
for a Salpeter IMF, $58 \%$ of the total cosmic amount of metals (all the elements heavier than He) is built. 
Spiral galaxies also represent important metal producers, contributing up to the $40 \%$ of the total metal budget, 
whereas irregulars represent negligible contributors, only with the $2 \%$. 
Of the total comoving density of metals, $\sim 60 \%$ is represented by oxygen,  which is the most abundant element 
after hydrogen and $^{4}$He.\\
ii) In the case of a Salpeter IMF, we predict a total metal comoving density  
$\rho_{Z}=9.37 \cdot 10^{6} M_{\odot} Mpc^{-3}$, smaller than the value inferred from cluster galaxies by Mushotzky \& Loewenstein 
(1997) of $\rho_{Z}=1.4 \cdot 10^{7} M_{\odot} Mpc^{-3}$. 
This difference is due to the fact that clusters represent biased (and not average) sites of metal production 
in the universe, for several  reasons. First,  they have formed in zones of high overdensity with respect to the field, 
thus experiencing an enhanced star formation and consequent metal production. 
Second, the galaxy proportions in clusters are not representative 
of the average galaxy population, since in clusters a very high percentage is represented by spheroids 
(up to the $80 \%$, Renzini et al. 1993), whereas studies based on the optical luminosity function 
indicate that $\sim 50 \%$ of the field galaxy population have discs (FHP98). 
For all these reasons, 
we think that metal production in clusters is 
not representative of the cosmic average.\\
iii) We have calculated the mean metal abundance in galaxies at the present time. 
Our results have provided an approximately solar value of $<Z>_{gal}= 0.0175$, confirming previous results by other authors based on 
different approaches (Edmunds \& Phillips 1997).\\
iv) We calculate the cosmic rate of $^4$He production for spheroids, discs and irregulars. 
We predict that in BCDs the ISM abundance of $^4$He, as due only to stars,   
should be between $\Delta Y=0.0013$ and $\Delta Y=0.002$ for a Scalo and 
a Salpeter IMF, respectively. 
Observations of the total $^4$He abundance in local BCD galaxies provide a present-time value of $Y=0.2456 \pm 0.012$ (Pagel et al. 1992).  
By subtracting to this value the $^4He$ produced by stars,  we can provide an independent estimate of the primordial 
abundance, namely  $Y_{p}=0.244 \pm 0.012$. Such an estimate is in good agreement with all the available appraisals of $Y_{p}$ 
based on $^{4}He$ detection in extragalactic $HII$ regions.\\ 
v) We find that the average $[O/Fe]_{*,E}$ ratio in stars 
in spheroids should be $+0.4$ dex, in very good agreement with the $[\alpha/Fe]$ values commonly measured in 
the dominant stellar populations of ellipticals (e. g., Thomas, Maraston \& Bender, 2002). 
On the other hand, our calculated abundance ratios of $\alpha$ to Fe elements 
in the gas of ellipticals are subsolar, with an average $[O/Fe]_{g,E} \sim -0.33$. 
This is consistent with recent XMM observations of the X-ray emitting gas in nearby ellipticals (Mushotzky 2002 
and references therein). The same ratios for spirals are $\sim +0.1$ dex for the stellar component and almost solar 
($\sim +0.01$ dex) for the gas component.\\
vi) We have performed a self consistent census of all the baryons and metals in the local universe, 
and compared it to different appraisals by other authors. We have found that 
the vast majority of the baryons lies outside galaxies in a diffuse warm IGM, but this is not true 
for the metals. The majority of metals (the 
$\sim 52 \% $ in our case) is locked up in stars, whereas the $\sim 9 \%$ is in the ISM in galaxies, with the remaining $ 39 \%$ 
ejected by spheroidal galaxies into the IGM. This is valid for $O$ and all the $\alpha$-elements, representing the bulk 
of the metals in the universe. A special case is represented by iron: we find that the $84 \%$ of all the Fe produced by 
spheroids (and exactly half of the total amount produced by all galaxies) is expelled into the IGM. 
This is a natural consequence of the galactic wind scenario, according to which ellipticals develop an 
early wind (in our case after a time between $0.3$ and 1 Gyr since their formation) 
emptying the galaxy of all the residual gas. 
At the occurrence of the wind, the bulk of the $\alpha$ elements 
 has already been produced and is locked up in stars. On the other hand, the production of the bulk of Fe has still to take place, 
since it occurs on longer timescales by means of type Ia SNe explosions.\\
vii) X-ray observations of the intra-cluster hot gas yield, for the Fe abundance of the ICM, values 
between $0.1$ and $\sim 1 $ solar. If we divide our predicted Fe density in the IGM by the average 
Fe abundance $X_{Fe,ICM}=0.3 X_{Fe,\odot}$ inferred for the intra-cluster gas by Renzini (1997),  
we obtain an estimate of the IGM baryon density of $\Omega_{b,IGM} \simeq  0.0126 $, which implies a 
a resulting total baryon density $\Omega_{b}=0.017$, i.e. lower 
than what current observations of CMB and of the deuterium abundance in QSO absorbers indicate ($\Omega_{b}=0.036-0.09$). 
This leads us to conclude that $X_{Fe,ICM} = 0.3 X_{Fe,\odot}$ represents an overestimation of the average Fe abundance of the IGM.  
In fact, by assuming $\Omega_{b}=0.08$ as the fiducial value for the total Baryonic density, our results imply a 
value of $\Omega_{IGM}=0.0753$ and an average IGM Fe abundance of $0.05$ solar. The main parameter in such a determination 
is the normalization of the local field luminosity function, uncertain by a factor of $2$.\\
Finally, it is worth recalling that 
all of our results are weakly dependent on the assumed redshift of formation for the whole galaxy population. 
This implies that from the present study it is impossible to derive information about the epoch of major galaxy 
formation. Of great interest in this direction would be the study of the cosmic evolution of the mean metal abundance 
of the universe, along with the evolution of the total baryonic/metal budget. For instance, from the amount of metals observed in Lyman 
Break galaxies at $z \sim 2.5 $ and assuming that they are protospheroids, in principle it would be possible to infer when the main epoch of 
spheroid formation has been, how many spheroids have 
already been observed and how many are still missing, 
either because still unformed or obscured by dust. 
The development of all of these topics is deferred to a forthcoming paper.

\section*{Acknowledgments}
We wish to thank Antonio Pipino for many helpful discussions.  
We are thankful to Cristina Chiappini and Laura Portinari for a careful reading of the manuscript and for 
several valuable comments. 
Finally, we wish to thank the anonymous referee for many suggestions that 
improved the quality of the paper.

%
%
%

\label{lastpage}

\end{document}